\documentclass[12pt]{article}
\pdfoutput=1
\usepackage{amsmath}
\usepackage{amssymb}
\usepackage{graphicx}
\usepackage{subcaption}
\usepackage{url}
\usepackage{arydshln}
\usepackage[sort&compress, numbers, merge]{natbib}

\newcommand{\be}{\begin{equation}}
\newcommand{\ee}{\end{equation}}

\newcommand{\bea}{\begin{eqnarray}}
\newcommand{\eea}{\end{eqnarray}}
\newcommand{\mat}{\begin{pmatrix}}
\newcommand{\rix}{\end{pmatrix}}

\renewcommand{\bar}{\overline}

\newcommand{\vv}{\mathbf}

\def\beq{\begin{equation}}
\def\eeq{\end{equation}}
\def\beqn{\begin{equation}}
\def\eeqn{\end{equation}}

\newcommand\iso[2]{\mbox{${}^{#2}${\rm #1}}}
\def\he#1{\iso{He}{#1}}

\usepackage[colorlinks=true, linkcolor=magenta, citecolor=blue,
urlcolor=black]{hyperref}

\DeclareMathOperator{\arccot}{arccot}

\begin{document}

\begin{titlepage}
\begin{center}

\hfill UT-16-33 \\
\vspace{2mm}
\hfill 	IPMU-16-0176 \\
\vspace{2mm}
\hfill \today

\vspace{3.0cm}
{\large \bf Effects of QCD bound states on dark matter relic abundance}

\vspace{1.0cm}
{\bf Seng Pei Liew}$^{(a)}$ and 
{\bf Feng Luo}$^{(b)}$ 
\vspace{1.0cm}

{\it
$^{(a)}${Department of Physics, The University of Tokyo, Tokyo 113-0033, Japan}\\
$^{(a)}${Kavli IPMU (WPI), UTIAS, The University of Tokyo, Kashiwa, Chiba 277-8583, Japan}\\
}

\vspace{1cm}
\abstract{We study scenarios where there exists an exotic massive particle charged under QCD in the early Universe. We calculate the formation and dissociation rates of bound states formed by pairs of these particles, and apply the results in dark matter (DM) coannihilation scenarios, including also the Sommerfeld effect. We find that on top of the Sommerfeld enhancement, bound-state effects can further significantly increase the largest possible DM masses which can give the observed DM relic abundance, by $\sim 30 - 100\%$ with respect to values obtained by considering the Sommerfeld effect only, for the color triplet or octet exotic particles we consider. In particular, it indicates that the Bino DM mass in the right-handed stop-Bino coannihilation scenario in the Minimal Supersymmetric extension of the Standard Model (MSSM) can reach $\sim 2.5$ TeV, even though the potential between the stop and antistop prior to forming a bound state is repulsive. 
We also apply the bound-state effects in the calculations of relic abundance of long-lived or metastable massive colored particles, and discuss the implications on the BBN constraints and the abundance of a super-weakly interacting DM. 
The corrections for the bound-state effect when the exotic massive colored particles also carry electric charges, and the collider bounds are also discussed.
}
\end{center}
\end{titlepage}

\setcounter{footnote}{0}

\tableofcontents

\section{Introduction}
In scenarios of physics beyond the Standard Model (SM), the early Universe may have been inhabited by exotic particles charged under QCD. Due to their strong interactions with SM particles, they were initially in thermal equilibrium and later froze out. 

If the colored particle is stable, stringent constraints for a strongly interacting dark matter (DM) particle apply (see~\cite{astro-ph/0705.4298} and references therein). On the other hand, if metastable, it may first freeze out and then decay. 
This can happen in, e.g., the $R$-parity conserving Minimal Supersymmetric extension of the Standard Model (MSSM) where the next-to-lightest supersymmetric particle (NLSP) is colored, and the lightest supersymmetric particle (LSP), which is a DM candidate, is extremely weakly interacting (the superWIMP), such as the gravitino or axino. Typically, observational and experimental bounds are applicable only to a metastable colored particle with a lifetime $\gtrsim 0.1 \, \text{sec}$. In particular, important constraints can be derived from Big-Bang nucleosynthesis (BBN) (see e.g.~\cite{astro-ph/0408426}). Moreover, the decays of the colored particles into superWIMPs can contribute non-thermally to the relic density of DM. 

The massive colored particle can also play a role in determining the relic abundance of the DM when the latter is a weakly interacting massive particle (WIMP), especially when the mass of the colored particle is almost degenerate with the DM. In this case, coannihilations can significantly reduce the relic abundance of the DM~\cite{Griest:1990kh}. This has been considered in the MSSM scenarios where the neutralino is the LSP, and the coannihilator is a squark (in particular a lighter top squark)~\cite{hep-ph/9911496,hep-ph/0112113,hep-ph/0301106,1104.3566,1111.4467,1212.5241,1409.2898,1404.5571,1412.7672,1501.03164} or a gluino~\cite{1412.7672,hep-ph/0402208,0903.5204,0905.1148,0908.0731,1011.1246,1011.5518,1310.0643,1403.0715,1408.5102,1503.07142,1504.00504,1510.03498}. The coannihilator can also be a Kaluza-Klein (KK) quark in the Universal Extra Dimension (UED) models~\cite{hep-ph/0206071,hep-ph/0509118,hep-ph/0509119,hep-ph/0605280,1012.2577,1611.06760}.

The main purpose of this paper is to study the effects of exotic massive colored particles on DM relic abundance, assuming that they share the same discrete symmetry stabilizing DM (e.g., $R$-parity in supersymmetric models and $KK$-parity in UED models).  We calculate the DM relic abundance in scenarios where the colored particle coannihilates with the WIMP. We also discuss implications of a metastable colored particle on BBN and the DM relic abundance in the superWIMP scenario. 

The major update of our work compared to previous calculations is the inclusion of the effects of QCD bound states of the exotic colored particles, $X_1$ and $X_2$. 
The formation of a QCD bound state, $\eta$, occurs via the process $X_1X_2\to \eta g$, in which the gluon, $g$, takes away the binding energy. 
$\eta$ can then decay to SM particles via the annihilation of its constituents, $X_1$ and $X_2$. The net effect is to remove $X_1$ and $X_2$ from the thermal bath. As will be shown in the following, these processes increase significantly the effective annihilation cross section of the $X$'s or DM. We note that due to color charge conservation, the QCD potential between $X_1$ and $X_2$ before they form a bound state can be different from the one after they become the constitutes of $\eta$, with a gluon emitted. Moreover, in contrary to the Sommerfeld effect~\cite{sommerfeld1931} which can enhance the effective annihilation cross section when the potential between the incoming $X_1$ and $X_2$ is attractive, bound-state effects can enhance the effective annihilation cross section no matter whether the potential between the incoming $X_1$ and $X_2$ is attractive, repulsive or zero. 
In addition, we calculate the corrections to the bound-state effects when the colored particle is electrically charged.  

Furthermore, we study the compatibility of these scenarios with collider constraints, especially those derived from the Large Hadron Collider (LHC) experiments. Studying LHC constraints is particularly timely as the undergoing LHC experiments are capable of excluding masses of colored particles up to the TeV scale. It is worth noting that while the DM relic abundance and BBN constraints often imply upper limits on the masses of colored particles or DM, collider results impose lower limits on the masses of colored particles. Studying both BBN/DM relic density and collider constraints therefore probe the experimentally allowed parameter ``window" of the long-lived colored particles or the coannihilating DM scenarios. 

The rest of the paper is organized as follows. In Section~\ref{sec:formalism}, we discuss the formation, dissociation and annihilation decay of bound states formed by a pair of massive colored particles in the early Universe, focusing in particular on the cases of complex scalars (S3) or Dirac fermions (F3) in the color $\text{SU(3)}$ fundamental representation, and real scalars (S8) or Majorana fermions (F8) in the adjoint representation. In Section~\ref{sec:relic}, we calculate the DM relic abundance in massive colored particle - WIMP coannihilation scenarios and the contributions from decays of metastable massive colored particles on the relic density of a superWIMP DM, by including the Sommerfeld and bound-state effects in the Boltzmann equation. We also consider the constraints from BBN on the long-lived massive colored particle scenarios, and discuss the electric charge corrections for the bound-state effect. The thermally-averaged $s$- and $p$-wave annihilation cross sections needed in the calculations are listed in Appendix~\ref{app:mat}. In Section~\ref{sec:collider}, we study the collider limits on the exotic massive colored particles. Finally, we summarize our conclusions in Section~\ref{sec:conclusion}. 

\section{Bound state formalism}
\label{sec:formalism}
In this section, we discuss the formation and dissociation of the QCD bound state, $X_1X_2\leftrightarrow \eta g$, in the early Universe. 
DM bound-state formation due to some new abelian gauge force was considered in~\cite{0905.3039}, and 
a systematic study of its effects on DM relic abundance was performed in~\cite{1407.7874}. 
Using the SM non-abelian color $\text{SU(3)}$ group, gluino bound-state effects on neutralino DM coannihilation in the MSSM have been investigated in~\cite{1503.07142}. 
A field-theoretic framework for the computations of bound-state effects was established in~\cite{1505.00109}. 
A formalism for bound states based on non-relativistic effective theories was given in~\cite{1602.08105,1609.00474}. 
Bound-state formation due to a Yukawa potential was studied in~\cite{1611.01394}. 
See e.g.~\cite{1403.1077,1406.2276,1604.01776,1607.00374} for other phenomenological implications of DM bound-state formation. 

We are concerned with massive colored particles with masses $m_{X_1}, m_{X_2} \gg \Lambda_{\rm QCD}$. In the early Universe before the quark-hadron phase transition era, the long-range interaction between two massive colored particles can be described by a perturbative one-gluon exchange Coulomb-like potential, which has the form:
\be
V(r) = - \, \frac{\zeta}{r} \, ,
\label{V(r)}
\ee
in which $\zeta$ is determined by the quadratic Casimir coefficients of the color representations of the individual colored particles, $X_1$ and $X_2$  ($C_{X_1}$ and $C_{X_2}$, respectively), as well as of the one by taking $X_1$ and $X_2$ together in a specific color state ($C_{X_1X_2}$):
\be
\zeta = \frac{1}{2}\left(C_{X_1} + C_{X_2} - C_{X_1X_2}\right) \alpha_s \, ,
\label{eq:color-factor}
\ee
where $\alpha_s > 0$ is the QCD coupling strength. A positive, negative or zero value of $\zeta$ gives an attractive, repulsive or zero potential, respectively. 

The colored particles we consider in this paper include a complex scalar and a Dirac fermion in the color $\text{SU(3)}$ fundamental representation, a real scalar and a Majorana fermion in the adjoint representation. The $X_1X_2$ combinations are $S3\bar{S3}$, $F3\bar{F3}$, $S8S8$ and $F8F8$, abbreviated in the following as S3, F3, S8 and F8, respectively, and hence $m_{X_1} = m_{X_2} \equiv m_X$. Examples of S3 and F8 are a squark-antisquark pair and a gluino-gluino pair, respectively, in the MSSM. A $KK$ quark-antiquark pair in models of UED is a realization of F3. One can also build models for the S8 case~\cite{1506.07110,1510.03434}.
The product of a color triplet and an anti-triplet is decomposed as 
\be
\vv{3}\otimes\vv{\bar{3}}=\vv{1}\oplus\vv{8},
\label{eq:33bardec}
\ee 
and the product of two color octets is decomposed as 
\be
\vv{8}\otimes\vv{8}=\vv{1}_S\oplus\vv{8}_A\oplus\vv{8}_S\oplus\vv{10}_A\oplus\vv{\bar{10}}_A\oplus\vv{27}_S,
\label{eq:88dec}
\ee
where the subscripts ``$S$" and ``$A$" indicate symmetric and anti-symmetric color states, respectively. Therefore, the relevant quadratic Casimir coefficients of the color representations for our calculations are $C_\vv{1} = 0$, $C_\vv{3} = 4/3$, $C_\vv{8} = 3$, $C_\vv{10} = 6$ and $C_\vv{27} = 8$. 

\footnote{We note that at a temperature $T$ of the Universe, the screening effect from the quarks and gluons in the thermal plasma induces a thermal mass $m_{th} \sim \sqrt{\alpha_s} T$ to the gluon, modifying the QCD Coulomb potential to a Yukawa one. However, as emphasized in~\cite{1611.01394}, the Coulomb potential is a good approximation as long as the momentum transfer between the two incoming particles, $\sim m_X \sqrt{\frac{T}{m_X}}$, is larger than $m_{th}$. We can see that this condition is well satisfied at the usual freeze-out temperature $T \sim m_X / 20$, and further better satisfied for the bound-state effect calculation since the effect of which is important at even lower temperature $T \sim \alpha_s^2 m_X$, as will be shown in the next section.}In principle, a bound state can form as long as the potential for it is attractive. In this paper, we focus on the color-singlet bound state, since it is expected to be the deepest bound (i.e., the ground state) and the most copiously produced one, in analogy to atomic physics~\footnote{In~\cite{1604.01776}, the formation of bound states at excited energy levels is discussed for bound-state effects in the late Universe, and it is found that the total bound-state formation cross section is dominated by levels with principle quantum numbers $n < \zeta / v_{rel}$, where $v_{rel}$ is the relative velocity of the incoming particles. Compared to the ground state, the contribution from the excited states enhances the total bound-state formation cross section by a logarithmic factor $ \sim {\rm log}(\zeta / v_{rel})$, which is significant for $v_{rel} \sim 10^{-3}$ in the galactic halo. However, in the early Universe at temperatures relevant for the dark matter relic abundance calculation, $v_{rel}$ is of order $10^{-1}$, so that $\zeta / v_{rel} \sim 1$. Moreover, compared to the ground state, the excited states are easier to be dissociated by gluons in the thermal bath, while the dissociation is not a concern for bound states in the late Universe. Therefore, the contribution from the excited states is not significant for the relic abundance calculation. Nevertheless, in the next section we will also show results with a factor of 2 enhancement of the bound-state effect from the considerations of the excited states contribution as well as other uncertainties in our calculations.}. Therefore, the coefficient, $\zeta$, in Eq.~(\ref{V(r)}) for the bound states in the S3 and F3 cases is $1/2 \times (4/3 + 4/3 - 0) \alpha_s = (4/3) \alpha_s$, and is $1/2 \times (3 + 3 - 0) \alpha_s = 3 \alpha_s$ for the S8 and F8 cases. In all the four cases, we consider that the color-singlet bound states have total orbital angular momentum $L = 0$ and spin $S = 0$~\footnote{The total wave function of the bound state is a product of the spatial, spin and color parts of the wave functions. For the S8 (F8) case, because of the nature of identical particles, the total wave function should be symmetric (anti-symmetric). $L = 0$ gives symmetric (symmetric) spatial wave function, and $S = 0$ gives symmetric (anti-symmetric) spin wave function. Together with the symmetric color wave function of the color-$\vv{1}_S$ state, indeed the requirement for the total wave function is satisfied.}. The normalized spatial wave function of such a bound state is 
\bea
\phi_{\eta}(r)=(\pi a^3)^{-1/2}e^{-r/a},
\label{eq:bsfi}
\eea
where $a$ is the Bohr radius, given as
\be
a = (\zeta \mu)^{-1},
\label{eq:bohr}
\ee
where $\mu \equiv m_{X_1} m_{X_2}/(m_{X_1}+m_{X_2}) = m_X/2$ is the reduced mass. The binding energy of the bound state is 
\be
E_B=\frac{\zeta^2 \mu}{2}.
\label{eq:binding}
\ee

\subsection{Bound-state formation, dissociation, and annihilation decay}
\label{sec:boundstate}
 
We start with a general description of the formation, dissociation and annihilation decay processes of bound states without specifying the color representation of the particles.

First of all, due to color charge conservation, the emission (absorption) of a gluon during bound-state formation (dissociation) makes the color representation of the bound state $\eta$ not necessarily be the same as the free pair $X_1X_2$. Therefore, the coefficient in the Coulomb potential for the free pair, denoted as $\zeta'$, is not necessarily equal to the one for the bound state. In particular, $\zeta'$ can be negative, so that the potential is repulsive for the free pair. As will be shown, at high temperature in the early Universe, the massive colored particles can have enough kinetic energy to overcome a repulsive potential to form a bound state. 

We follow~\cite{1503.07142} to calculate the bound-state formation and dissociation cross sections, where the method is adapted from the calculations of the photoelectric effect for an atom~\cite{Berestetskii:1971}. The essence of the calculation is to evaluate the transition matrix element between the bound state and the free pair state:
\be
\mathcal{M}_{fi}=\int \phi^*_f(-i\frac{\vec{\nabla}\cdot\vec{\epsilon}^c}{\mu})e^{i\vec{k}\cdot \vec{r}}\phi_i d^3\vec{r},
\label{eq:Mfi}
\ee
where $\phi_f$ is the wave function of the free pair and $\phi_i \equiv \phi_\eta (r)$. The gluon has a momentum $\vec{k}$ and a polarization $\vec{\epsilon}^c$, where ``c'' is the color index. 

For the free pair, the normalized spatial part of the wave function is (see Section 136 of~\cite{Landau:1991wop})
\bea
\phi_f=\frac{1}{2|\vec{p}|} \sum_{L=0}^{\infty} i^L(2L+1)e^{-i\delta_L}R_{pL}(r)P_L(\frac{\vec{p}\cdot \vec{r}}{|\vec{p}|r}),
\label{eq:free}
\eea    
where $|\vec{p}|$ is the relative momentum of the free pair, expressed in terms of the reduced mass and their relative velocity as $|\vec{p}| = \mu v_{rel}$. $P_L(\frac{\vec{p}\cdot \vec{r}}{|\vec{p}|r})$ is the Legendre polynomial and $\delta_L$ (a real number) is the phase shift. Note that the form of the radial function $R_{pL} (r)$ for an attractive potential between the free pair is different from the one for a repulsive potential (see Section 36 of~\cite{Landau:1991wop} for details). Consider the spatial part of the wave functions only, the differential dissociation cross section is given as 
\be
d\sigma^0_{dis}=\alpha_s\frac{\mu |\vec{p}|}{2\pi\omega}|\mathcal{M}_{fi}|^2 d\Omega_{\vec{p}},
\label{eq:ddis}
\ee
where $\omega \equiv |\vec{k}|$ is the energy of the gluon. The explicit $\alpha_s$ factor in the above equation comes from the coupling between the emitted gluon and the massive colored particle. In the Lagrangian of the quantum field theory, this coupling is from the covariant derivative of the kinetic term of the massive colored particle, and it takes the form of $i g_s {\bf T}_c$, where $g_s = \sqrt{4 \pi \alpha_s}$ is the strong coupling, and ${\bf T}_c$ are the generator matrices for the color representation in which the massive colored particle lies. We will specify ${\bf T}_c$ in the next subsection when we consider the color part of the wave functions for the four cases of our interest. 

We use the dipole approximation~\footnote{We refer the reader to~\cite{1503.07142} for details. For the four cases of our interest (S3, F3, S8, F8), we have checked that the dipole approximation is always justified. Also, the kinetic energy of the bound state is negligible compared to the gluon energy, so that $\omega \approx E_B + {1\over2} \mu v_{rel}^2$.}, $e^{i\vec{k}\cdot \vec{r}}\simeq 1$, to calculate the transition matrix element Eq.~(\ref{eq:Mfi}). This means that only the $L=1$ term in $\phi_f$ has a non-zero contribution. Also, considering that it is the absolute square of the transition matrix element that appears in Eq.~(\ref{eq:ddis}), we can drop the phase factors and rewrite $\phi_f$ as 
\be
\phi_f = {3 \over 2 |\vec{p}|} R_{p1} (r) P_1 ({\vec{p} \cdot \vec{r} \over |\vec{p}| r}) \, . 
\label{eq:freestatewavefunction} 
\ee

Defining dimensionless quantities
\be
\nu \equiv |\zeta'|/v_{rel}
\label{eq:v}
\ee
and
\be
\kappa \equiv \zeta/|\zeta'|, 
\label{eq:kappa}
\ee 
we can write down the integrated dissociation cross section, averaged over the incoming gluon spin polarizations. The result depends on whether the free pair feels an attractive (denoted by the subscript ``$a$'') or a repulsive (denoted by the subscript ``$r$'') Coulomb potential:
\bea
\sigma^{0}_{dis,a} &=& {2^{9} \pi^2 \over 3} \alpha_{s} a^2 \left(\frac{E_B}{\omega}\right)^4\frac{1+\nu^{2}}{1+(\kappa\nu)^2} {e^{-4 \nu \arccot(\kappa\nu)} \over 1-e^{-2 \pi \nu}} \kappa^{-1}\, , 
\label{eq:dis0a} 
\\
\sigma^0_{dis,r} &=& {2^{9} \pi^2 \over 3} \alpha_{s} a^2 \left(\frac{E_B}{\omega}\right)^4\frac{1+\nu^{2}}{1+(\kappa\nu)^2} {e^{4 \nu \arccot(\kappa\nu)-2\pi \nu} \over 1-e^{-2 \pi \nu}} \kappa^{-1}\, .
\label{eq:dis0r}
\eea
In the case that the free pair feels no potential (denoted by the subscript ``$free$'', and see Section 33 of~\cite{Landau:1991wop} for the radial function), we find
\bea
\sigma^0_{dis,free} &=& {2^{9} \pi^2 \over 3} \alpha_{s} a^2 \left(\frac{E_B}{\omega}\right)^4 \frac{(a \mu v_{rel})^3}{2\pi \left[1 + (a \mu v_{rel})^2 \right]} \, .
\label{eq:dis0free}
\eea
One can check that in the $\zeta' \rightarrow 0$ limit, Eq.~(\ref{eq:dis0a}) and Eq.~(\ref{eq:dis0r}) both become Eq.~(\ref{eq:dis0free}). 

The superscript ``0'' in the above three equations indicates that we have considered the spatial part of the wave functions only, while the full wave functions are products of spatial, color and spin wave functions. Also, if the particles are identical, one needs to symmetrize or anti-symmetrize the wave functions. The full dissociation cross section, $\sigma_{dis}$, after taking into account color, spin and the symmetry factors, is related to the bound-state formation cross section, $\sigma_{bsf}$, via the Milne relation:
\bea
\sigma_{bsf}=\frac{g_{\eta}g_g\omega^2}{g_{X_1}g_{X_2}\left(\mu v_{rel}\right)^2}\sigma_{dis}, 
\label{eq:bsf}
\eea
where $g_{g, X_1, X_2, \eta}$ are the degrees of freedom of gluon, $X_1$, $X_2$ and $\eta$, respectively. Note that if $X_1$ and $X_2$ are identical, the left-hand side of Eq.~(\ref{eq:bsf}) has to be multiplied by 1/2 to avoid double counting the number of bound-state formation reactions. 

Bound state can be destroyed not only by the dissociation process, but also by decays. Moreover, the decays can happen in two ways: the constituent particles inside the bound state can annihilate between themselves (annihilation decay) or an individual constituent particle can decay by itself. The effects of these two kinds of decays on the relic density of the metastable colored particles or DM are different. Since we assume that the constituent particles in the bound state have the same discrete symmetry as the DM particle, the annihilation decay to SM particles removes, for example, two $R$-odd numbers in supersymmetry, while the individual constitute particle decay does not change the $R$-odd number. For the colored particle coannihilating in the WIMP DM scenario and the metastable colored particle in the superWIMP DM scenario, the individual constituent particle decay rate is suppressed either by the small mass difference or by the very small coupling between the massive colored particle and the DM particle~\footnote{This is the case in the MSSM for a Bino-like neutralino LSP coannihilating with a stop, when the two-body decay of the stop into top and neutralino is kinematically forbidden, and indeed coannihilation is responsible for giving the correct DM relic abundance for the small mass difference range~\cite{1404.5571}. For a neutralino LSP coannihilating with a gluino, the gluino decay rate can be very suppressed by the small mass difference as well as by large squark masses in the propagator~\cite{1503.07142}. For a gravitino or axino LSP, its coupling with the NLSP is suppressed by the Planck or the Peccei-Quinn scale.}, while the annihilation decay rate is not suppressed and is proportional to the large mass of the colored particle. Therefore, we will hereafter neglect the individual constitute particle decay rate compared to the annihilation decay rate. 

\subsection{Results for S3, F3, S8 and F8}
Here we present the full bound-state formation and dissociation cross sections for the cases of S3, F3, S8 and F8, as well as the annihilation decay rates. 
\\

\noindent {\bf S3 and F3}

Since we consider that the bound state is a color-singlet state, the emission (absorption) of a gluon in the bound-state formation (dissociation) process dictates that for both S3 and F3 the free pair state must be in a color-octet state (see Eq.~(\ref{eq:33bardec})), due to color charge conservation. The normalized color wave function is $\delta_{kj}/\sqrt{3}$ for the bound state, and $\lambda^b_{ij}/\sqrt{2}$ for the free pair, where $\lambda^b_{ij}$ are the Gell-Mann matrices, and the color indices $i,j,k=1-3,\, b=1-8$. The generator ${\bf T}_c$ takes the form $\lambda^c_{ki}/2\,$. Therefore, the color part of the wave functions contributes to $\sigma_{dis}$ as
\be
\left|\frac{\lambda^b_{ij}}{\sqrt{2}}\frac{\lambda^c_{ki}}{2}\frac{\delta_{kj}}{\sqrt{3}}\right|^2= \left|\frac{\delta^{bc}}{\sqrt{6}}\right|^2
=\frac{4}{3} \, .
\ee

For S3, there is no spin wave function to worry about. While for F3, without considering the bound state, a pair of heavy colored fermion and anti-fermion can have $3/4$ chance in spin-triplet configurations with $S=1$ and $1/4$ chance in a spin-singlet configuration with $S=0$. Since we only consider a bound state with $S=0$, then by neglecting the spin-orbit interaction we will only consider a free pair that is also in $S=0$ state. Therefore, we consider that both the bound state and the free pair have the same spin wave function, given as
\be
(\uparrow \downarrow - \downarrow \uparrow)/\sqrt{2} \,,
\ee
so that the spin part of the wave functions does not introduce a factor for $\sigma_{dis}$. However, in the next section we will see that when including the bound-state formation and dissociation cross sections in the Boltzmann equation, we need to introduce an additional factor of $1/4$ for F3 compared to S3 to take into account the fact that we have only considered the $S=0$ possibility in the former. 

Putting the factor of $1/8$ from the incoming gluon color averaging, the factor of $4/3$ from the color part of the wave functions, and by noticing that the free pair has a repulsive potential with $\zeta' = 1/2 \times (4/3 + 4/3 - 3) \alpha_s= (-1/6) \alpha_s$ (see Eq.~(\ref{eq:color-factor})), we get the full dissociation cross section for S3 and F3, 
\bea
\sigma_{dis}^{S3,F3} &=& {1 \over 8} \times {4 \over 3} \times \sigma^{0}_{dis,r} \, ,
\eea
in which the quantities inside $\sigma^{0}_{dis,r}$ are given in Eqs.~(\ref{eq:bohr}), (\ref{eq:binding}), (\ref{eq:v}) and (\ref{eq:kappa}) with $\zeta = (4/3) \alpha_s$. From Eq.~(\ref{eq:bsf}), the bound-state formation cross sections are
\bea
\sigma_{bsf}^{S3} &=& \frac{1 \times 16}{3 \times 3} \frac{\omega^2}{\left(\mu v_{rel}\right)^2} \times \sigma_{dis}^{S3,F3} \, 
\eea
and
\bea
\sigma_{bsf}^{F3} &=& \frac{1 \times 16}{6 \times 6} \frac{\omega^2}{\left(\mu v_{rel}\right)^2} \times \sigma_{dis}^{S3,F3} \, ,
\eea
where the degrees of freedom are written explicitly. 

For the bound-state annihilation decay, we consider the dominant decay mode only, which is the two-gluon final state (see e.g.~\cite{1103.3503}), and the results are 
\bea
\Gamma^{S3}_\eta &=& \frac{1}{3}\mu\alpha_s^2 \zeta^3
\label{eq:anndecays3}
\eea
and
\bea
\Gamma^{F3}_\eta &=& \frac{2}{3}\mu\alpha_s^2 \zeta^3, 
\label{eq:anndecayf3}
\eea
where $\zeta = (4/3)\alpha_s$. In the above two equations, the $\alpha_s$ factor explicitly written is evaluated at the scale of $2 m_X$, while the $\alpha_s$ inside $\zeta$ is evaluated at the scale of the inverse Bohr radius, $a^{-1}$.
\\

\noindent {\bf S8 and F8}

Due to the nature of identical particles, the total wave functions need to be symmetric for S8 whereas anti-symmetric for F8. 

The F8 case was studied in detail in~\cite{1503.07142}, and the result for the gluon dissociation of a color-$\vv{1}_S$ bound state with ($S=0, L=0$) into a free pair in an $\vv{8}_A$ state with ($S=0, L=1$) is 
\bea
\sigma_{dis}^{F8} &=& 3 \times 4 \times {1 \over 8} \times {1 \over 2} \times \sigma^{0}_{dis,a} \, ,
\eea
where the factor $3$ comes from the color part of the wave functions together with the generator ${\bf T}_c = -i f_{cde}$ in the coupling between the gluon and the massive colored particle, where $f_{cde}$ are the SU(3) structure constants. The factor $4$ comes from symmetrization of the spatial part of the bound-state wave function ($L=0$) and anti-symmetrization of the spatial part of the free pair wave function ($L=1$). $1/8$ comes from the color averaging of the incoming gluon. The factor $1/2$ is introduced to avoid double counting of the two identical massive colored particles in the outgoing free pair phase-space integration. The spin part of the wave functions do not introduce any extra factor. The quantities inside $\sigma^{0}_{dis,a}$ are given in Eqs.~(\ref{eq:bohr}), (\ref{eq:binding}), (\ref{eq:v}) and (\ref{eq:kappa}) with $\zeta = 3 \alpha_s$ and $\zeta' = 1/2 \times (3 + 3 - 3) \alpha_s= (3/2) \alpha_s$. 

The S8 case is exactly the same as the F8 case, namely, a transition from the color-$\vv{1}_S$ bound state with ($S=0, L=0$) into the $\vv{8}_A$ free pair state with ($S=0, L=1$). The only difference is that while for the F8 case the $S=0$ state means that the spin wave function is anti-symmetric (i.e., a spin-singlet configuration), for the S8 case there is no spin to worry about, so that for the latter the total wave functions for both bound state and free pair state are symmetric, as they should be. Therefore, we have
\bea
\sigma_{dis}^{S8} &=& \sigma_{dis}^{F8} \; \; \equiv \; \; \sigma_{dis}^{S8,F8} \, . 
\eea
The corresponding bound-state formation cross sections are
\bea
\sigma_{bsf}^{S8} &=& 2 \times \frac{1 \times 16}{8 \times 8} \frac{\omega^2}{\left(\mu v_{rel}\right)^2} \times \sigma_{dis}^{S8,F8} \, 
\label{eq:bsfs8}
\eea
and
\bea
\sigma_{bsf}^{F8} &=& 2 \times \frac{1 \times 16}{16 \times 16} \frac{\omega^2}{\left(\mu v_{rel}\right)^2} \times \sigma_{dis}^{S8,F8} \, ,
\label{eq:bsff8}
\eea
where the factor of $2$ in the front of the right-hand side of the two equations is actually the factor of $1/2$ which would have been in the left-hand side of Eq.~(\ref{eq:bsf}) to avoid double counting the number of bound-state formation reactions. In Eqs.~(\ref{eq:bsfs8}) and (\ref{eq:bsff8}), the degrees of freedom are written explicitly. 

Again, we consider the dominant two-gluon annihilation decay channel only, and the annihilation decay rates are (see e.g.~\cite{1103.3503}), 
\bea
\Gamma^{S8}_\eta &=& \frac{9}{4}\mu\alpha_s^2 \zeta^3
\label{eq:anndecays8}
\eea
and
\bea
\Gamma^{F8}_\eta &=& \frac{9}{2}\mu\alpha_s^2 \zeta^3, 
\label{eq:anndecayf8}
\eea
where $\zeta = 3 \alpha_s$. The scales of evaluating the $\alpha_s$ explicitly written and the one inside $\zeta$ in the above two equations are understood similarly as in the S3 and F3 cases. 

\subsection{Thermal averaging}

To study the bound-state effects on the relic abundance of the massive colored particles or the DM, we need the thermally-averaged bound-state dissociation and annihilation decay rates, as well as the formation cross section times the relative velocity of the free pair, since it is these quantities that appear in the Boltzmann equation which determines the evolution of the density of the massive colored particles or the DM with temperature, $T$. By defining two dimensionless variables, $z \equiv E_B / T$ and $u \equiv \frac{1}{2} \mu v_{rel}^2 /T$, we can rewrite $\sigma_{dis}$ and $\sigma_{bsf}$ as functions of $z$ and $u$, together with factors not changing with $T$. In particular, the relevant quantities are expressed as
\bea
\omega &=& E_B \left(1+ \frac{u}{z}\right) \, , 
\label{eq:omega} \\
\nu &=& \left(\frac{z}{u}\right)^{\frac{1}{2}} \kappa^{-1} \, ,\;\; \text{for} \;\; \zeta'\neq 0 \, , 
\label{eq:nu} \\
v_{rel} &=& \zeta \left(\frac{u}{z}\right)^{\frac{1}{2}} \, .
\label{eq:vrel}
\eea
The thermally-averaged bound-state dissociation rate is 
\be
\langle\Gamma\rangle_{dis} \;\;=\;\; g_g \frac{4 \pi}{\left( 2 \pi \right)^3} \int_{E_B}^\infty \sigma_{dis} \frac{\omega^2 d\omega}{e^{\omega/T} -1} \;\; = \;\; g_g \frac{4 \pi}{\left( 2 \pi \right)^3} \int_0^\infty \sigma_{dis} \frac{E_B^3 \left( 1+ \frac{u}{z}\right)^2 du}{z \left( e^{z+u} - 1\right)} \, .
\label{eq:disaverage}
\ee
The thermally-averaged bound-state formation cross section times relative velocity is 
\be
\langle \sigma v \rangle_{bsf} \;\; = \;\; \int_0^\infty \sigma_{bsf} v_{rel} f(v_{rel}) \left( 1+ \frac{1}{e^{\omega/T}-1}\right) d v_{rel} \, ,
\label{eq:thermalbsf}
\ee
where $f(v_{rel})$ is the Maxwell-Boltzmann distribution function for $v_{rel}$, given as 
\be
f(v_{rel})  =  \left(\mu \over 2 \pi T \right)^{3/2} 4 \pi v_{rel}^2 \, e^{-\frac{\mu v_{rel}^2}{2T}} \, .
\label{eq:velocitydis}
\ee
The factor $\frac{1}{e^{\omega/T}-1}$ in Eq.~(\ref{eq:thermalbsf}) accounts for the stimulated emission due to the gluons in the thermal bath. 
Using Eqs.~(\ref{eq:vrel}) and (\ref{eq:binding}), $\langle \sigma v \rangle_{bsf}$ can be rewritten as
\be
\langle \sigma v \rangle_{bsf} \;\; = \;\; \int_0^\infty \sigma_{bsf} \zeta \left(\frac{u}{z}\right)^{\frac{1}{2}} \frac{2}{\sqrt{\pi}} u^{1/2} e^{-u} \left( 1+ \frac{1}{e^{z+u} -1}\right) du  \, .
\ee
The thermally-averaged bound-state annihilation decay rate is 
\be
\langle \Gamma \rangle_\eta \;=\; \Gamma_\eta \langle \frac{m_\eta} {E_\eta} \rangle \; \approx \; \Gamma_\eta  \frac{\int_{0}^\infty \frac{m_\eta}{E_\eta} e^{-E_\eta/T} d^3 \vec{p}_\eta}{\int_{0}^\infty e^{-E_\eta/T} d^3 \vec{p}_\eta} \;=\; \Gamma_\eta \frac{K_1 (m_\eta / T)}{K_2 (m_\eta / T)} \, ,
\ee
where $m_\eta$ is the mass of the bound state, given as $m_\eta = m_{X_1} + m_{X_2} - E_B$. In the above formula, we have assumed the Maxwell-Boltzmann approximation for the bound-state equilibrium distribution, and $K_{1,2} (m_\eta / T)$ are the modified Bessel functions of the second kind. At $m_\eta \gg T$, $\langle \Gamma \rangle_\eta \approx \Gamma_\eta$. 

\section{Relic abundance calculation}
\label{sec:relic}
We need to set up a Boltzmann equation (or a coupled set of Boltzmann equations) to solve for the relic abundance of the massive colored particles in superWIMP scenario or of the DM in WIMP scenario. The necessary ingredients for the bound state in the Boltzmann equation, i.e., the thermally-averaged bound-state formation cross section, dissociation and annihilation decay rates, are given in the previous section. There is another important ingredient -- the Sommerfeld effect -- needs to be considered for massive annihilating particles feeling a long-range force. The Sommerfeld effect in the calculations of thermal relic abundance has been studied extensively in the literature (see e.g.~\cite{hep-ph/9806361,hep-ph/0610249,0706.4071,1005.4678,1008.2905,1010.2172,deSimone:2014pda,1601.04718}).  
Let us first briefly describe the Sommerfeld corrections to the annihilation cross sections of the massive colored particles of our interest.  

\subsection{Sommerfeld effect}
We consider the Sommerfeld effect resulting from the long-range Coulomb potential between the two   annihilating massive colored particles. 
Under the influence of a Coulomb potential of the form $V (r)= -\alpha/r$, the Sommerfeld corrected $s$-wave annihilation cross section can be written as
\be
\sigma v_{rel} \;\;=\;\;  a S(\alpha / v_{rel}) \, ,
\label{eq:somswave}
\ee
where
\be
S(\alpha / v_{rel}) = \frac{2 \pi \alpha / v_{rel}}{1-e^{-2 \pi \alpha / v_{rel}}} \, .
\ee
An attractive potential ($\alpha > 0$) results in an enhancement ($S>1$), while a repulsive potential ($\alpha < 0$) results in a suppression ($S<1$). In Eq.~(\ref{eq:somswave}), the perturbative $s$-wave cross section, $a$, does not depend on temperature. Therefore, the thermally-averaged Sommerfeld corrected $s$-wave cross section is $a \langle S (\alpha / v_{rel}) \rangle$, where  
\be
\langle S (\alpha / v_{rel}) \rangle = \int_0^\infty S (\alpha / v_{rel}) f (v_{rel}) dv_{rel} \, ,
\ee
with $f (v_{rel})$ given in Eq.~(\ref{eq:velocitydis}). 

For the massive colored particles of our interest, we expect that the dominate annihilation channels are $S3\bar{S3}$, $F3\bar{F3}$, $S8S8$ and $F8F8$ annihilation into a pair of gluons, $gg$, and into quark-antiquark pairs, $q\bar{q}$. We consider the Sommerfeld corrected $s$-wave cross sections and the tree-level $p$-wave cross sections in our calculation. The relevant expressions are collected in Appendix~\ref{app:mat}. To consider the Sommerfeld effect, we need to decompose an $s$-wave cross section into partial cross sections contributed from each of the two-body states in different color representations, as given in Eqs.~(\ref{eq:33bardec}) and (\ref{eq:88dec}), because different representations correspond to different Coulomb-like potentials. We follow the decompositions given in~\cite{deSimone:2014pda}. The thermally-averaged Sommerfeld factors are
\bea
\frac{\langle \sigma v_{rel} (S3\bar{S3} \; \text{or} \; F3\bar{F3} \to gg) \rangle_{s\text{-wave, Sommerfeld}}}{\langle \sigma v_{rel} (S3\bar{S3} \; \text{or} \; F3\bar{F3} \to gg) \rangle_{s\text{-wave, pertubative}}} &=& \frac{2}{7} \langle S(\frac{4\alpha_s/3}{v_{rel}}) \rangle+\frac{5}{7} \langle S(\frac{-\alpha_s/6}{v_{rel}}) \rangle  \, , \nonumber \\
\frac{\langle \sigma v_{rel} (F3\bar{F3}  \to q\bar{q}) \rangle_{s\text{-wave, Sommerfeld}}}{\langle \sigma  v_{rel} (F3\bar{F3}  \to q\bar{q}) \rangle_{s\text{-wave, pertubative}}} &=& \langle S(\frac{-\alpha_s/6}{v_{rel}}) \rangle \, ,\nonumber \\
\frac{\langle \sigma v_{rel} (S8S8 \; \text{or} \; F8F8 \to gg)  \rangle_{s\text{-wave, Sommerfeld}}}{\langle \sigma v_{rel} (S8S8 \; \text{or} \; F8F8 \to gg) \rangle_{s\text{-wave, pertubative}}} &=& \frac{1}{6} \langle S(\frac{3\alpha_s}{v_{rel}}) \rangle+\frac{1}{3} \langle S(\frac{3\alpha_s/2}{v_{rel}}) \rangle +\frac{1}{2} \langle S(\frac{-\alpha_s}{v_{rel}}) \rangle  \, , \nonumber \\
\frac{\langle \sigma v_{rel} (F8F8  \to q\bar{q}) \rangle_{s\text{-wave, Sommerfeld}}}{\langle \sigma v_{rel} (F8F8  \to q\bar{q}) \rangle_{s\text{-wave, pertubative}}} &=& \langle S(\frac{3\alpha_s/2}{v_{rel}}) \rangle \, .
\label{eq:somfactor}
\eea
The $s$-wave cross sections vanish for $S3\bar{S3}  \to q\bar{q}$ and $S8S8  \to q\bar{q}$. 

\subsection{Boltzmann equation}
\label{sec:boltzmann}
We now have all the ingredients to write down and solve the Boltzmann equation. The general formulae for $N$ species of exotic particles in calculating the thermal relic densities are given in~\cite{1510.03498}. In that formalism, the possibility of when the rates for interconverting some of the species are not sufficiently large is taken into account, and in such case a coupled set of Boltzmann equations, rather than a single Boltzmann equation, is needed. Also, a simple method for implementing the bound-state effects in the Boltzmann equations is given, and a detailed example for the gluino-gluino bound state is shown in~\cite{1503.07142}. 

In this paper, we consider two exotic species -- the massive colored particle and the DM, and by considering that they share the same discrete symmetry stabilizing DM, we assume that the decay of one massive colored particle produces one DM together with some SM particles. In the superWIMP scenario, because the DM has long been out of the thermal bath when the freeze-out of the massive colored particle happens, we only need to solve the Boltzmann equation for the latter. In the WIMP scenario, we are interested in the coannihilation between the two exotic species. As highlighted  in~\cite{1503.07142,1504.00504}, coannihilation is effective only if the interconversion rate between the two species is sufficiently large compared to the Hubble expansion rate, otherwise the two species would freeze out separately. Without committing to specific particle theory models, in this work we assume the interconversion rate is sufficiently large so that the two species freeze out together and we can use a single Boltzmann equation to calculate the DM density. We emphasize that this condition needs to be checked when one consider coannihilations in specific DM models. 

Also, as we are mainly interested in QCD interactions and further model specifications would have been needed, in our calculation we neglect the (co)annihilation cross sections for DM - DM and DM - massive colored particle, in comparing to the annihilation cross sections between the massive colored particles. In the WIMP DM scenario, indeed this is usually a good approximation. For example, for the right-handed stop-Bino coannihilation in the MSSM, the stop - antistop annihilation to $gg$ channel dominates the effective annihilation cross section in the coannihilation region for an $\mathcal{O}$(TeV) Bino~\cite{1501.03164}, if the ratio of the lighter stop - heavier stop - Higgs coupling to heavier stop mass is not very large~\cite{1404.5571}; for the gluino-neutralino coannihilation, the gluino pair annihilations to $gg$ and $q\bar{q}$ dominate over the neutralino-gluino and neutralino pair (co)annihilation cross sections. 
 
With these assumptions and approximations, in the WIMP DM scenario the evolution of the total $yield$, that is, the total number density of exotic particles (i.e., the DM $\chi$ and the massive colored particles $X$) over the entropy density, $\tilde{Y}\equiv Y_X + Y_\chi$, where $Y_X \equiv n_X/s$ and $Y_\chi \equiv n_\chi/s$, can be described by a single Boltzmann equation as follows:
\bea
\frac{d\tilde{Y}}{dx}=-\frac{xs}{H(m_\chi)}\left(1+\frac{T}{3g_{\ast s}}\frac{dg_{\ast s}}{dT}\right)\langle \sigma_{eff} v\rangle\left(\tilde{Y}^2-\tilde{Y}^2_{eq}\right),
\label{eq:bol}
\eea
where
\bea
x \equiv {m_\chi \over T}, \, s = {2 \pi^2 \over 45} g_{\ast s} T^3, \; H(m_\chi) \equiv H(T) x^2 = \left({4 \pi^3 G_N g_\ast \over 45}\right)^{1 \over 2} m_\chi^2 \, , 
\label{eq:defxsh}
\eea
and $G_N$ is the gravitational constant, $m_\chi$ is the DM mass, $g_{\ast s}$ and $g_\ast$ are the numbers of effectively massless degrees of freedom associated with the entropy density and the energy density, respectively. $\tilde{Y}_{eq}$ is the equilibrium value of the total yield, given as $\tilde{Y}_{eq} = Y_X^{eq} + Y_\chi^{eq} = (n_X^{eq} + n_\chi^{eq})/s$, where 
\be
n_{X}^{eq} = \frac{T}{2 \pi^2} g_{X} m_{X}^2 K_2 (m_{X}/T) \,, \;\; n_{\chi}^{eq} = \frac{T}{2 \pi^2} g_{\chi} m_{\chi}^2 K_2 (m_{\chi}/T)
\ee 
are the thermal equilibrium number densities~\cite{Srednicki:1988ce,Gondolo:1990dk,Edsjo:1997bg}.  $g_\chi$ is the DM degrees of freedom. $g_X$ is the degrees of freedom of massive colored particles, being 6, 12, 8 and 16 for the cases of S3, F3, S8 and F8, respectively. We note that for the S3 and F3 cases $n_X$ and $n_X^{eq}$ take into account both the massive colored particles and anti-particles, assuming that there is no asymmetry between the number densities of them. The condition of a sufficiently large interconversion rate between $\chi$ and $X$ comparing to the Hubble expansion rate guarantees that $n_X / n_\chi \approx n_X^{eq} / n_\chi^{eq}$, to a very good approximation~\footnote{By setting up a coupled set of Boltzmann equations for $\chi$ and $X$ with sufficiently large interconversion rate between them (through decay, inverse decay and scatterings with SM particles), one can check that the solutions of the set of Boltzmann equations, $n_\chi$ and $n_X$, satisfy this relation very well.}. 
By further defining 
\be
\Delta\equiv (m_X-m_\chi)/m_\chi \, , \;\; g_{eff}\equiv g_\chi+g_X(1+\Delta)^{3/2}e^{-\Delta x} \, ,
\ee 
the thermally-averaged effective annihilation cross section, $\langle \sigma_{eff} v\rangle$, can be written as
\bea
\langle \sigma_{eff} v\rangle=\langle \sigma v\rangle_{XX \to SM} \frac{g_X^2(1+\Delta)^3e^{-2\Delta x}}{g_{eff}^2} \, ,
\label{eq:sigmaeff}
\eea
where $\langle \sigma v\rangle_{XX \to SM}$ is the thermally-averaged total annihilation cross section of $X$'s into SM particles, with the Sommerfeld and bound-state effects taken into account, given as
\be
\langle \sigma v\rangle_{XX \to SM} = \langle \sigma v_{rel} (XX \to gg, q\bar{q})\rangle + \langle \sigma v \rangle_{bsf}\frac{\langle\Gamma \rangle_\eta}{\langle\Gamma \rangle_\eta+\langle\Gamma \rangle_{dis}} \, .
\label{eq:xxtosm}
\ee

For the S3 and F3 cases, since $n_X$ includes both particles and anti-particles, we need to time a factor of $1/2$ to the $\langle \sigma v\rangle_{XX \to SM}$ term in the right-hand side of Eq.~(\ref{eq:sigmaeff}), if we are using the usual spin and color averaged cross sections, as explained in the Appendix of~\cite{Srednicki:1988ce}. Furthermore, for the F3 case, we need to time another factor of $1/4$ to the $\langle \sigma v \rangle_{bsf}$ and $\langle\Gamma \rangle_{dis}$ in Eq.~(\ref{eq:xxtosm}), since we only consider the bound-state formation and dissociation for the $S=0$ state, which occurs with $1/4$ of the possibilities of the total numbers of the spin configurations for two spin-$1/2$ fermions. 

The physical interpretation of the above formulae is as follows. First of all, since the mass of the bound state, $m_\eta = 2 m_X - E_B$, is much larger than $m_X$ (note that $E_B \sim \mathcal{O}(10^{-2}) \, m_X$ for $\alpha_s \sim 0.1$), the  equilibrium number density of the bound state is negligible compared to the one for $X$, for $m_X \gg T$. Also, since the bound-state annihilation decay rate, $\langle\Gamma \rangle_\eta$, is many orders of magnitude larger than the Hubble expansion rate for a massive colored particle with a mass of $\mathcal{O}(1 - 100)$ TeV (note that $\langle\Gamma \rangle_\eta / H \sim \alpha_s^5 G_N^{-1/2} m_X T^{-2} > \alpha_s^5 G_N^{-1/2} m_X^{-1}$, for $m_X > T$), which is of our interest in this paper, the yield of the bound state keeps decreasing and is always negligible compared to the one for $X$. These explain why we do not have to worry about the yield of the bound state in the Boltzmann equation~(\ref{eq:bol}). 

The second term on the right-hand side of Eq.~(\ref{eq:xxtosm}) accounts for the bound-state effect: once formed, the bound-state can be destroyed either by dissociation back to individual $X$'s or by annihilation decay to SM particles (as explained in Sec.~\ref{sec:boundstate}, we neglect the individual $X$ decay rate compared to the annihilation decay rate); only the latter process reduces the total number of $X$'s. At high temperatures, a large fraction of the gluons in the thermal bath have enough energy to break up the bound states before they can annihilation decay, so that the bound-state effect is negligible. With the temperature falling, this energetic fraction becomes smaller, and eventually when the temperature becomes comparable and then is below the bound-state binding energy, the annihilation decay process dominates over the dissociation process, and the bound-state effect helps to reduce the total number of $X$'s. Since we assume that $X$ and the DM particle share the same discrete symmetry which stabilizes DM, the net effect is to reduce the DM relic abundance. In other words, the formation and the subsequent annihilation decay of bound states enhance as a new channel of the annihilation of $X$'s. We note that to have the bound-state effect reduce the total number of $X$'s efficiently, the bound-state formation rate should be larger or comparable to the Hubble expansion rate, otherwise bound states essentially cannot form in the first place. Indeed eventually the bound-state effect will cease to be effective as the building block of the bound state, $n_X$, is decreasing with the expanding of the Universe, as long as the bound state formation cross section does not increase too fast with the decrease of temperature so that it compensates the decreasing of $n_X$. 

Following the usual procedure to obtain the DM relic abundance~\cite{Srednicki:1988ce,Gondolo:1990dk,Edsjo:1997bg} by integrating Eq.~(\ref{eq:bol}) from a small value of $x$ (when $\tilde{Y} = \tilde{Y}_{eq}$) to the current temperature of the Universe (when essentially $x \to \infty$), we get $\tilde{Y}_0$, where the subscript ``$0$'' indicates today, and the DM relic abundance is 
\be
\Omega_\chi h^2 = 2.755 \times 10^8 \frac{m_\chi}{\text {GeV}} \tilde{Y}_0 \, .
\label{eq:relic}
\ee

The above formulae can describe the evolution of the yield of the massive colored particles in the superWIMP scenario by the following modifications: in Eqs.~(\ref{eq:bol}), (\ref{eq:defxsh}) and (\ref{eq:relic}), change $\tilde{Y}$ to $Y_X$, $\tilde{Y}_{eq}$ to $Y_X^{eq}$, $m_\chi$ to $m_X$, and $\langle \sigma_{eff} v\rangle$ to $\langle \sigma v\rangle_{XX \to SM}$, which is given in Eq.~(\ref{eq:xxtosm}). After solving for the relic abundance of $X$ would have today, if it had not decayed, the amount of DM produced from the decays of $X$'s is obtained as 
\be
\Omega_{\text{SW}}^\text{non-th} h^2=\frac{m_{\text{SW}}}{m_{X}}\Omega_{X}h^2 \, ,
\label{eq:relicsw}
\ee  
where $m_{\text{SW}}$ is the mass of the superWIMP DM, and the superscript ``non-th'' indicates that there can be also thermally produced amount, $\Omega_{\text{SW}}^\text{th} h^2$, which contributes to the total DM relic abundance together with $\Omega_{\text{SW}}^\text{non-th} h^2$. 

\subsection{Bound-state effects}
\label{sec:boundeffect}

\begin{figure}
\begin{center}
\begin{tabular}{c c}
\hspace{-0.3cm}
\includegraphics[height=5.5cm]{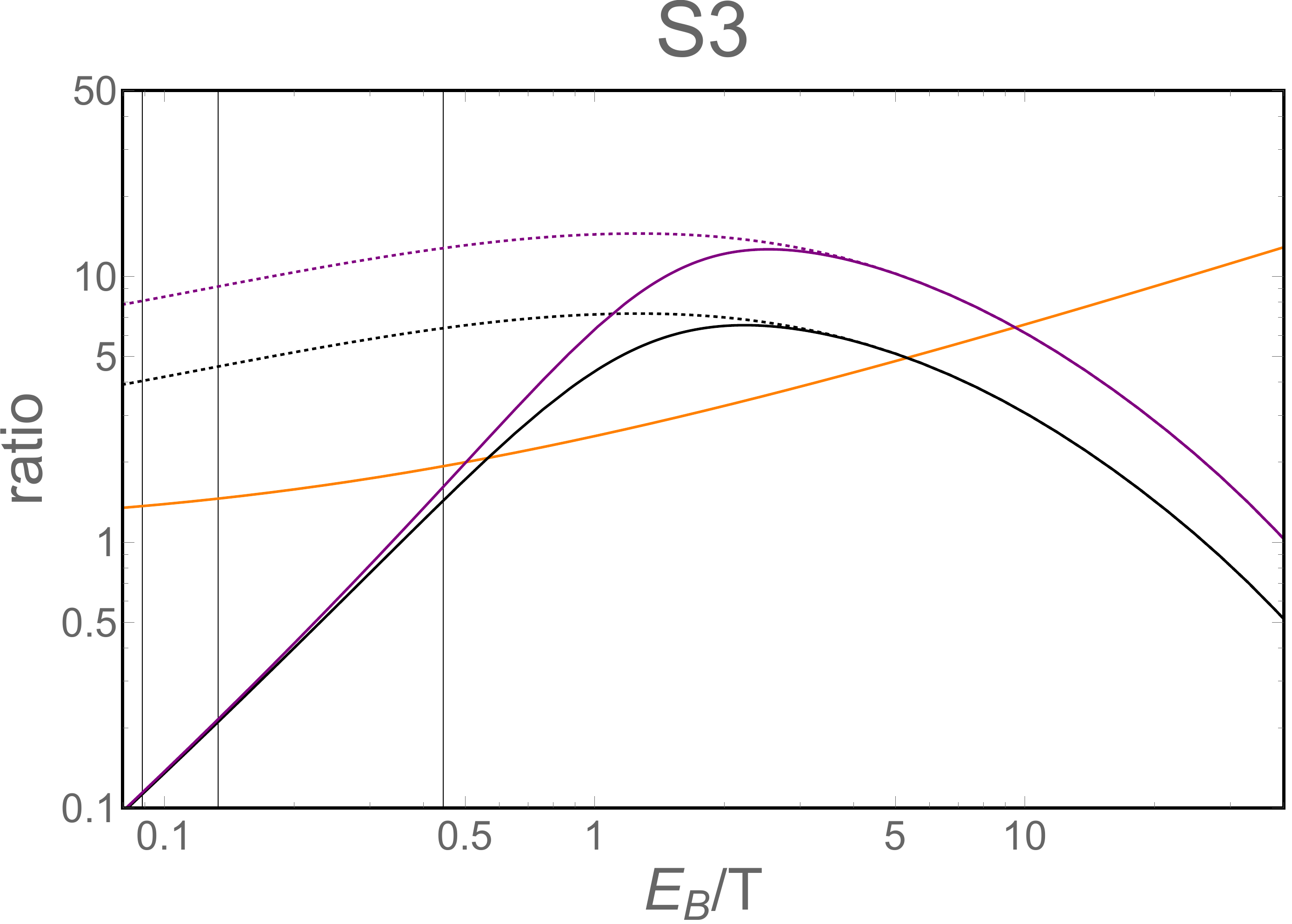} & 
\hspace{-0.3cm}
\includegraphics[height=5.5cm]{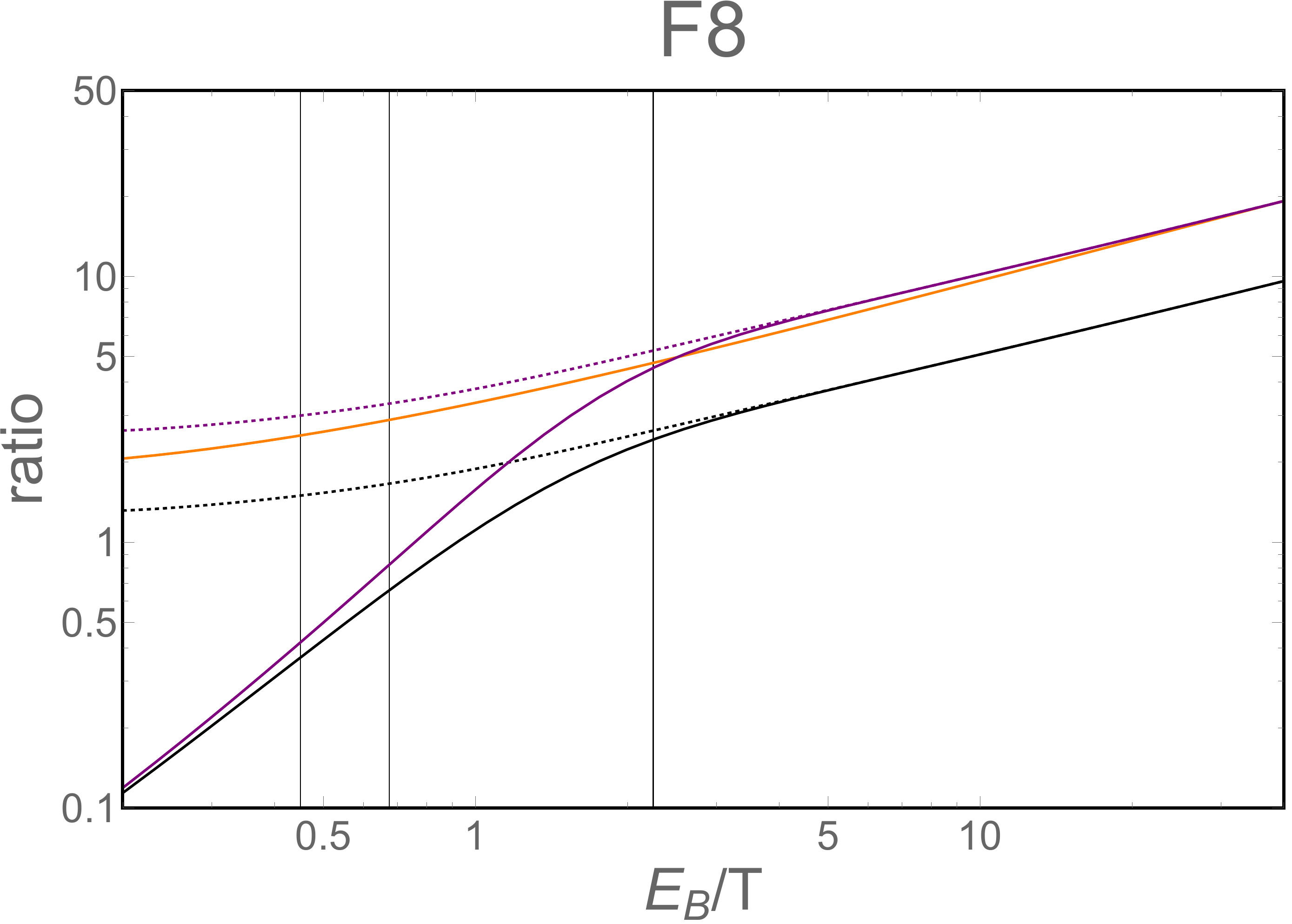} \\
\end{tabular}
\end{center}   
\caption{\label{fig:ratio}\it
The thermally-averaged Sommerfeld corrected $XX \to gg, q\bar{q}$ annihilation cross section (orange curve), bound-state formation cross section with (solid black curve) and without (dotted black curve) considering bound state dissociation and annihilation decay rates, as functions of $E_B/T$, for the S3 (left panel) and F8 (right panel) cases. The purple curves have the same meaning as the corresponding black curves, but multiplied by a factor of 2 in the bound-state formation cross section (solid and dotted) and dissociation rate (solid). All curves are normalized to the tree-level thermally-averaged $XX \to gg, q\bar{q}$ annihilation cross section without Sommerfeld correction. The thin vertical black lines correspond to, from left to right, $m_X/T=20,30,100$.
}
\end{figure}

To get an idea of the size of bound-state effects, we plot in Fig.~\ref{fig:ratio} with orange curves the ratio of thermally-averaged Sommerfeld corrected $XX \to gg, q\bar{q}$ annihilation cross section to the one without Sommerfeld correction, i.e., 
\be
\frac{\langle \sigma v_{rel} (XX \to gg, q\bar{q})\rangle_{\text{Sommerfeld}} }{\langle \sigma v_{rel} (XX \to gg, q\bar{q})\rangle_{\text{w/o Sommerfeld}}} \, ,
\ee 
the thermally-averaged bound-state formation cross section with (solid black curve) and without (dotted black curve) considering bound state dissociation and annihilation decay rates, also normalized to the tree-level $XX \to gg, q\bar{q}$ annihilation cross section, i.e., 
\be
\frac{\langle \sigma v \rangle_{bsf}\frac{\langle\Gamma \rangle_\eta}{\langle\Gamma \rangle_\eta+\langle\Gamma \rangle_{dis}}}{\langle \sigma v_{rel} (XX \to gg, q\bar{q})\rangle_{\text{w/o Sommerfeld}}} 
\;\;\; \text{and} \;\;\; 
\frac{\langle \sigma v \rangle_{bsf}}{\langle \sigma v_{rel} (XX \to gg, q\bar{q})\rangle_{\text{w/o Sommerfeld}}} \, ,
\ee
respectively, and with purple curves for the same ones but multiplied by a factor of 2 in the bound-state formation cross section (solid and dotted) and dissociation rate (solid), that is~\footnote{To account for the errors involving the evaluations of $\alpha_s$'s~\cite{0807.0211}, the corrections from a more accurate QCD potential and the thermal mass of the gluon~\cite{deSimone:2014pda}, as well as the possibility that excited bound states may also contribute to the bound-state effect~\cite{0905.3039}, we plot purple curves in this and following figures with an uncertainty of a factor of 2.},
\be
\frac{2 \langle \sigma v \rangle_{bsf}\frac{\langle\Gamma \rangle_\eta}{\langle\Gamma \rangle_\eta+ 2 \langle\Gamma \rangle_{dis}}}{\langle \sigma v_{rel} (XX \to gg, q\bar{q})\rangle_{\text{w/o Sommerfeld}}} 
\;\;\; \text{and} \;\;\; 
\frac{2 \langle \sigma v \rangle_{bsf}}{\langle \sigma v_{rel} (XX \to gg, q\bar{q})\rangle_{\text{w/o Sommerfeld}}} \, .
\ee
The left and right panels are for the S3 and F8 cases, respectively~\footnote{We use a common value, $\alpha_s = 0.1$, for all the $\alpha_s$'s appearing in the formulae for the curves in Fig.~{\ref{fig:ratio}} and the upper left, upper right and lower left panels in Fig.~\ref{fig:elecorrratio}. In this way, all the ratios do not depend on $m_X$. However, we note that in other parts of this paper and other plots, $\alpha_s$'s are evaluated differently: the $\alpha_s$'s appearing in the Sommerfeld factors in Eq.~(\ref{eq:somfactor}) are evaluated at $\beta m_X$ which is the typical scale of the momentum transfer of the soft-gluon exchanges which are responsible for the Sommerfeld effect~\cite{hep-ph/9806361}, and we take $\beta = 0.3$, which is roughly the average of the thermal velocities of the $X$'s at the freeze-out temperature; the $\alpha_s$'s in the bound-state formation cross sections and dissociation rates, as well as in the $\zeta$ part of the annihilation decay rates (given in Eqs.~(\ref{eq:anndecays3}), (\ref{eq:anndecayf3}), (\ref{eq:anndecays8}) and (\ref{eq:anndecayf8})), are evaluated at the bound-state inverse Bohr radius scale; the $\alpha_s$'s in the tree-level $XX \to gg, q\bar{q}$ annihilation cross sections and the ones appearing explicitly in the bound-state annihilation decay rates, are evaluated at the scale of $2 m_X$.}. 
As can be observed from both panels, at early times when $E_B /T \ll 1$, the solid black curves are much smaller than 1, although the dotted black curves have already bigger than 1. This is because for this high temperature range, a large fraction of the gluons in the thermal bath are energetic enough to break up the newly formed bound states before the latter can annihilation decay, that is, $\langle\Gamma \rangle_{dis} \gg \langle\Gamma \rangle_\eta$. Also, a factor of 2 increase of the bound-state formation cross section and also in the dissociation rate (since they are related via the Milne relation) has little effect. With the decrease of temperature, $\langle\Gamma \rangle_{dis}$ becomes smaller and eventually negligible compared to $\langle\Gamma \rangle_\eta$ when $E_B/T \gg 1$, so that the solid and dotted black (and purple) curves merge. By comparing the orange and solid black (and purple) curves, one can see that when $E_B/T \gtrsim 1$, the size of the full-fledged bound-state effect is of the same order and even larger than that of the Sommerfeld enhancement. 
 
There is a qualitative difference between the S3 and F8 cases at $E_B/T \gg 1$: while in the F8 case the solid black and purple curves keep growing with the decrease of temperature, in the S3 case they are decreasing after achieving maximum values around $E_B/T \sim 2$. This is due to the fact that an S3 incoming pair feels a repulsive potential prior to forming a bound state, while the potential is attractive for an F8 incoming pair. At lower temperature, there is a smaller fraction of the incoming S3 pairs which have enough kinetic energy to overcome the repulsive potential to form bound states. While for the F8 case, a low temperature (and hence small velocities) favors the formation of bound states, similar to that of the Sommerfeld enhancement. 

\begin{figure}
\begin{center}
\begin{tabular}{c c}
\hspace{-0.3cm}
\includegraphics[height=5.5cm]{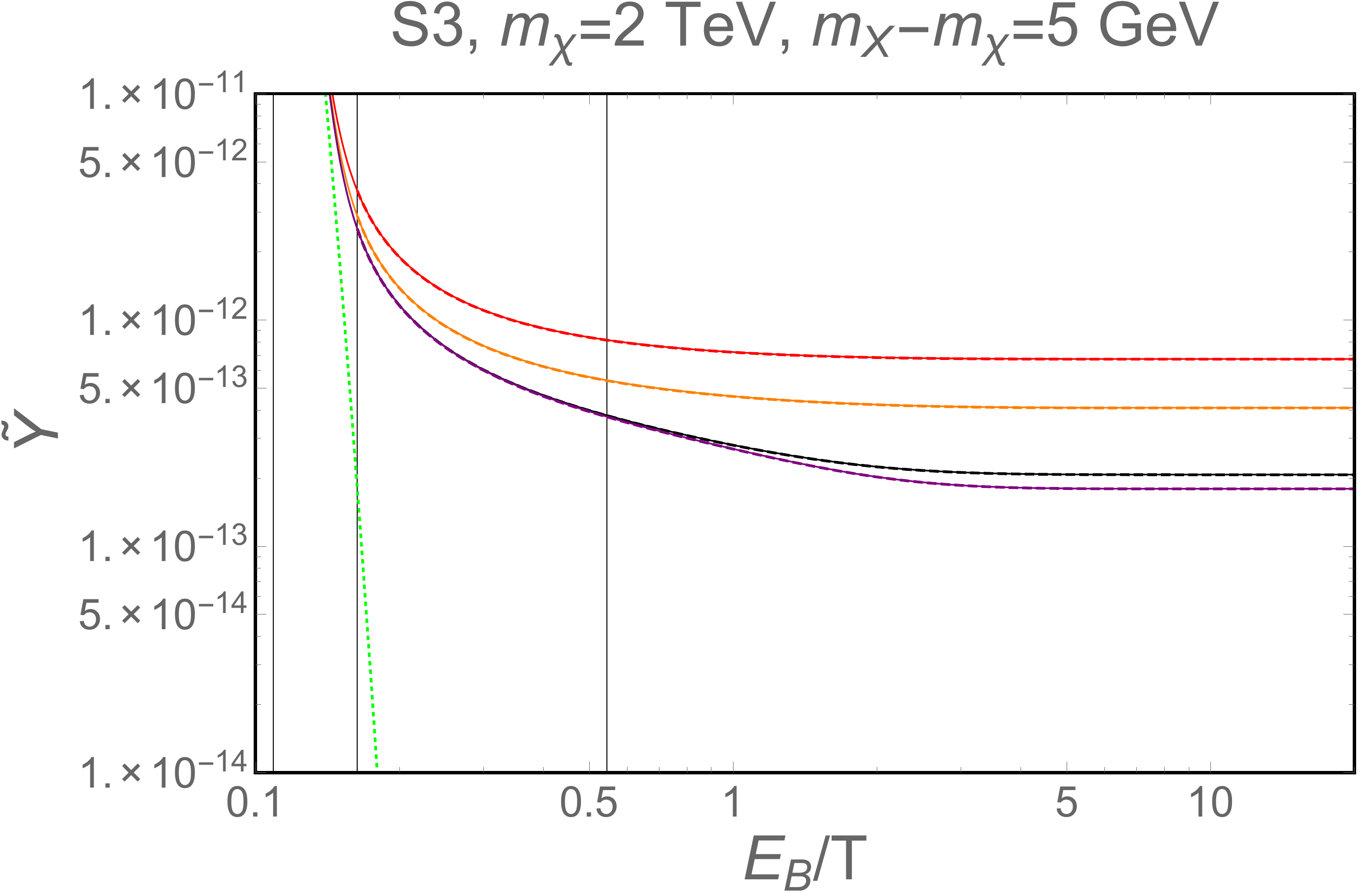} & 
\hspace{-0.3cm}
\includegraphics[height=5.5cm]{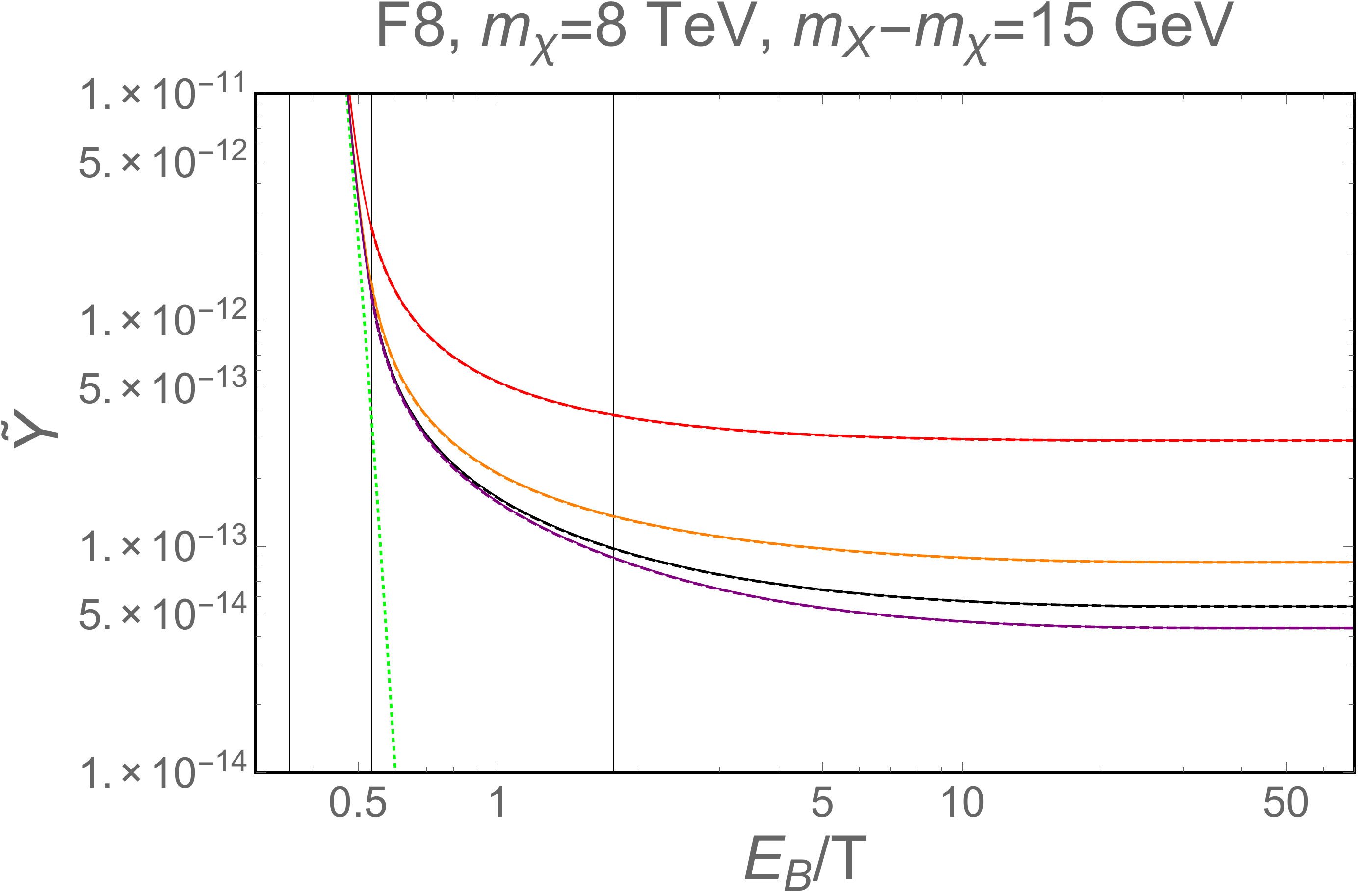} \\
\end{tabular}
\end{center}   
\caption{\label{fig:yieldvsz}\it
The yield $\tilde{Y}$ as a function of $E_B/T$ for S3 (left panel) and F8 (right panel) coannihilating with a WIMP DM. The solid red, orange, black and purple curves are results calculated from Eq.~(\ref{eq:bol}) without the Sommerfeld and bound-state effects, with the Sommerfeld effect but without bound-state effect, with both the Sommerfeld and bound-state effects, and with the Sommerfeld effect and a factor of 2 enlargement of the bound-state effect, respectively. The dashed red, orange, black and purple curves are the corresponding results given by the approximation Eq.~(\ref{eq:yapprox}). The green dotted line is for the equilibrium value $\tilde{Y}_{eq}$. The thin vertical black lines correspond to, from left to right, $m_X/T=20,30,100$.
}
\end{figure}

In order to see the bound-state effect on the evolution of the yield and on the result of final DM relic abundance, we plot in Fig.~\ref{fig:yieldvsz} the change of $\tilde{Y}$ with $E_B/T$ in the WIMP coannihilation scenarios for the S3 (left panel) and F8 (right panel) cases, by assuming a WIMP DM with the number of degrees of freedom $g_\chi = 2$. For S3 (F8), we take $m_\chi= 2$ TeV ($8$ TeV), and the mass difference between DM and the colored particle 5 GeV (15 GeV). The solid red, orange, black and purple curves are results calculated from Eq.~(\ref{eq:bol}) without the Sommerfeld and bound-state effects, with the Sommerfeld effect but without bound-state effect, with both the Sommerfeld and bound-state effect, and with Sommerfeld effect and a factor of 2 enlargement of the bound-state effect (i.e., $\langle \sigma v \rangle_{bsf} \to 2 \langle \sigma v \rangle_{bsf}$ and $\langle\Gamma \rangle_{dis} \to 2 \langle\Gamma \rangle_{dis}$), respectively. In particular, the limiting values of the solid black curves at $E_B/T \to \infty$ give roughly the correct DM relic abundance consistent with the observational value, as we will show more in the next subsection by parameter scans. The equilibrium yield, $\tilde{Y}_{eq}$, is shown by a green dotted line. The thin vertical black lines correspond to, from left to right, $m_X/T=20,30,100$. As can be seen from the plots, the thermal freeze-out, defined when $(\tilde{Y}/\tilde{Y}_{eq} -1)$ becomes order unity, happens between $m_X/T \sim 20$ and $ \sim 30$~\footnote{$m_X/T \approx m_\chi/T$ since $m_X - m_\chi \ll m_\chi$.}. After freeze-out, $\tilde{Y}$ keeps decreasing to its limiting value at $E_B/T \to \infty$, which can be more than one order of magnitude smaller than its value at freeze-out, especially when considering the Sommerfeld and bound-state effects. At $E_B/T \ll 1$, the black and purple curves are very close to the orange curve, indicating that comparing to the Sommerfeld effect, bound-state effect is not important for that temperature range. However, at $E_B/T \sim 1$, the black and purple curves depart from the orange curve, and eventually result in significantly lower values of $\tilde{Y}$ at $E_B/T \to \infty$ (the limiting values for the purple and black curves are $\sim 40\% - 60\%$ of the ones for the orange curve, for the two parameter sets shown). We can also see that the purple curves almost overlap with the black curves until $E_B/T \sim 1$, indicating that enlarge bound-state formation and dissociation by a factor of 2 has little effect on the yield at high temperature. These features are expected from the relative sizes of the Sommerfeld and bound-state enhancement factors illustrated in Fig.~\ref{fig:ratio}~\footnote{The relative sizes of the Sommerfeld and bound-state effects, as well as the positions of the vertical lines for $m_X/T = 20, 30, 100$ in Fig.~\ref{fig:ratio}, are a little different from the ones in Fig.~\ref{fig:yieldvsz}, due to the different $\alpha_s$'s used as mentioned earlier.}.

To understand the evolution of $\tilde{Y}$ more transparently, we also plot in Fig.~\ref{fig:yieldvsz} approximate solutions of $\tilde{Y}$ with dashed red, orange, black and purple curves by using the fact that $\tilde{Y}_{eq}$ becomes negligible compared to $\tilde{Y}$ shortly after freeze-out, so that Eq.~(\ref{eq:bol}) can take the approximate form
\bea
\frac{d\tilde{Y}}{dx} \approx -\frac{xs}{H(m_\chi)}\left(1+\frac{T}{3g_{\ast s}}\frac{dg_{\ast s}}{dT}\right)\langle \sigma_{eff} v\rangle \tilde{Y}^2 \,,
\eea
which has the solution
\bea
\tilde{Y}(x_2) = \left[\frac{1}{\tilde{Y}(x_1)}+\int^{x_2}_{x_1} \frac{xs}{H(m_\chi)}\left(1+\frac{T}{3g_{\ast s}}\frac{dg_{\ast s}}{dT}\right)\langle \sigma_{eff} v\rangle dx \right]^{-1} \, .
\label{eq:yapprox}
\eea
We plot the dashed curves by using Eq.~(\ref{eq:bol}) until $x_1 = 30$, at which $\tilde{Y}$ is already much larger than $\tilde{Y}_{eq}$, then we input the value of $\tilde{Y} (x_1 = 30)$ in Eq.~(\ref{eq:yapprox}) and use it to plot $\tilde{Y}$ at larger values of $x$. One can see that this approximation is very accurate, as the dashed curves completely overlap with the corresponding solid curves. It is easier to see from Eq.~(\ref{eq:yapprox}) that with a large $\langle \sigma_{eff} v\rangle$ for at least a range of not-too-large $x$ (in other words, when $n_X$ has not been too diluted away by Hubble expansion), the contribution from the integration part can be large compared to $1/\tilde{Y}(x_1)$. This explains why for the S3 case the bound-state effect on the result of the final relic abundance is still significant, even though the bound-state formation cross section decreases to zero at $x \to \infty$. 

\subsection{Coannihilations in the WIMP DM scenario}

\begin{figure}
\begin{center}
\begin{tabular}{c c}
\hspace{-0.3cm}
\includegraphics[height=8cm]{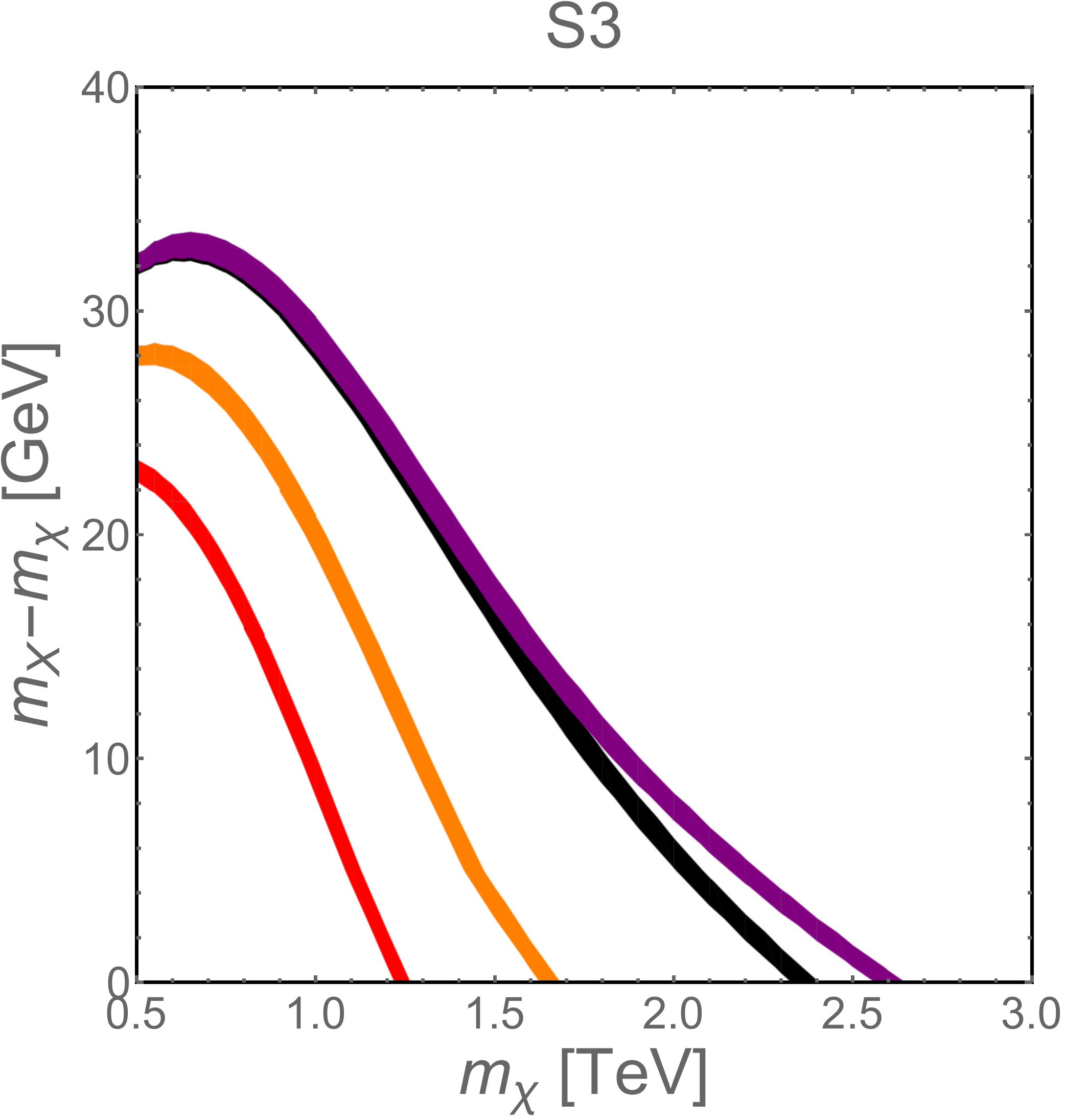} & 
\hspace{-0.3cm}
\includegraphics[height=8cm]{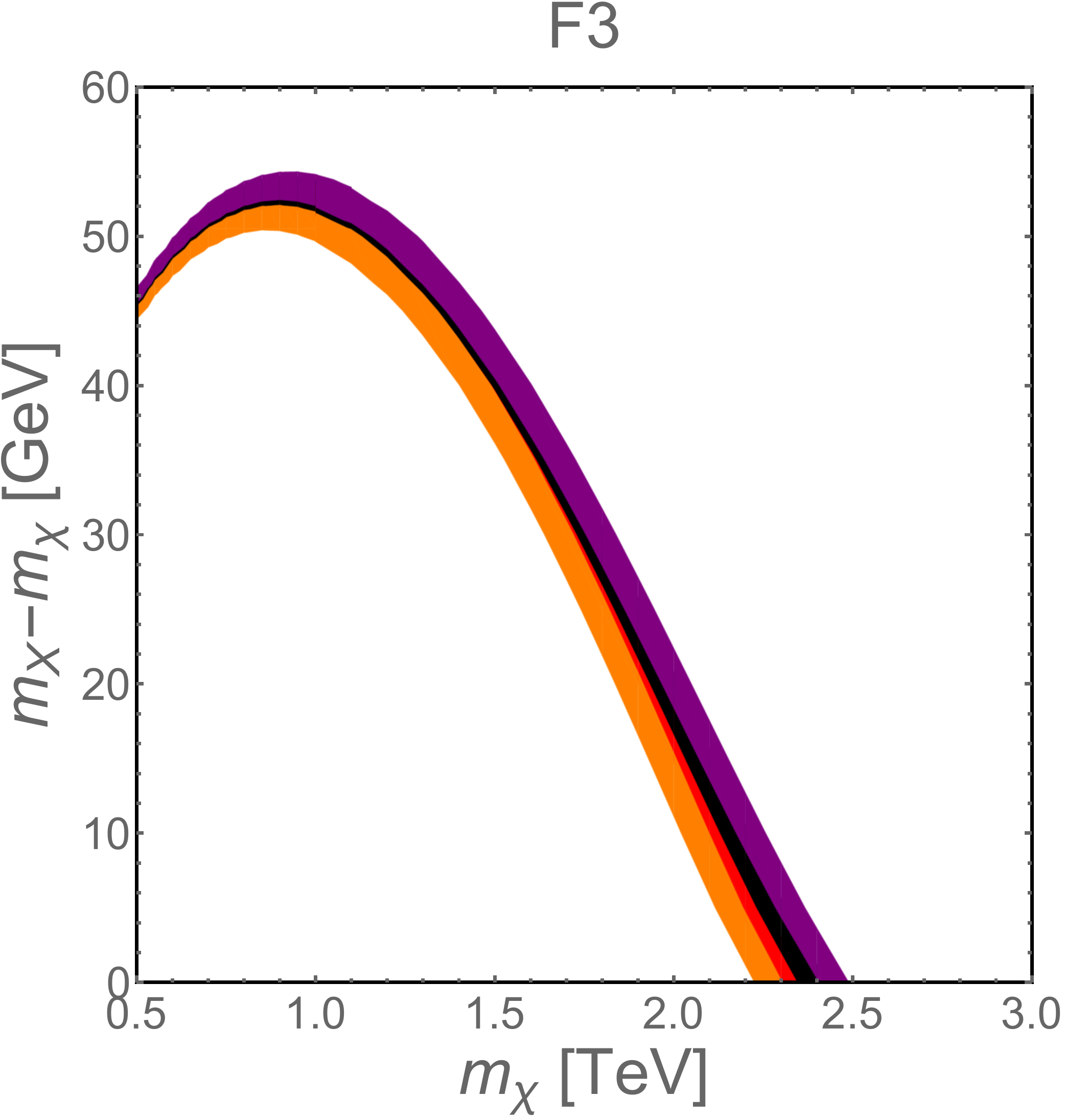} \\
\hspace{-0.3cm}
\includegraphics[height=8cm]{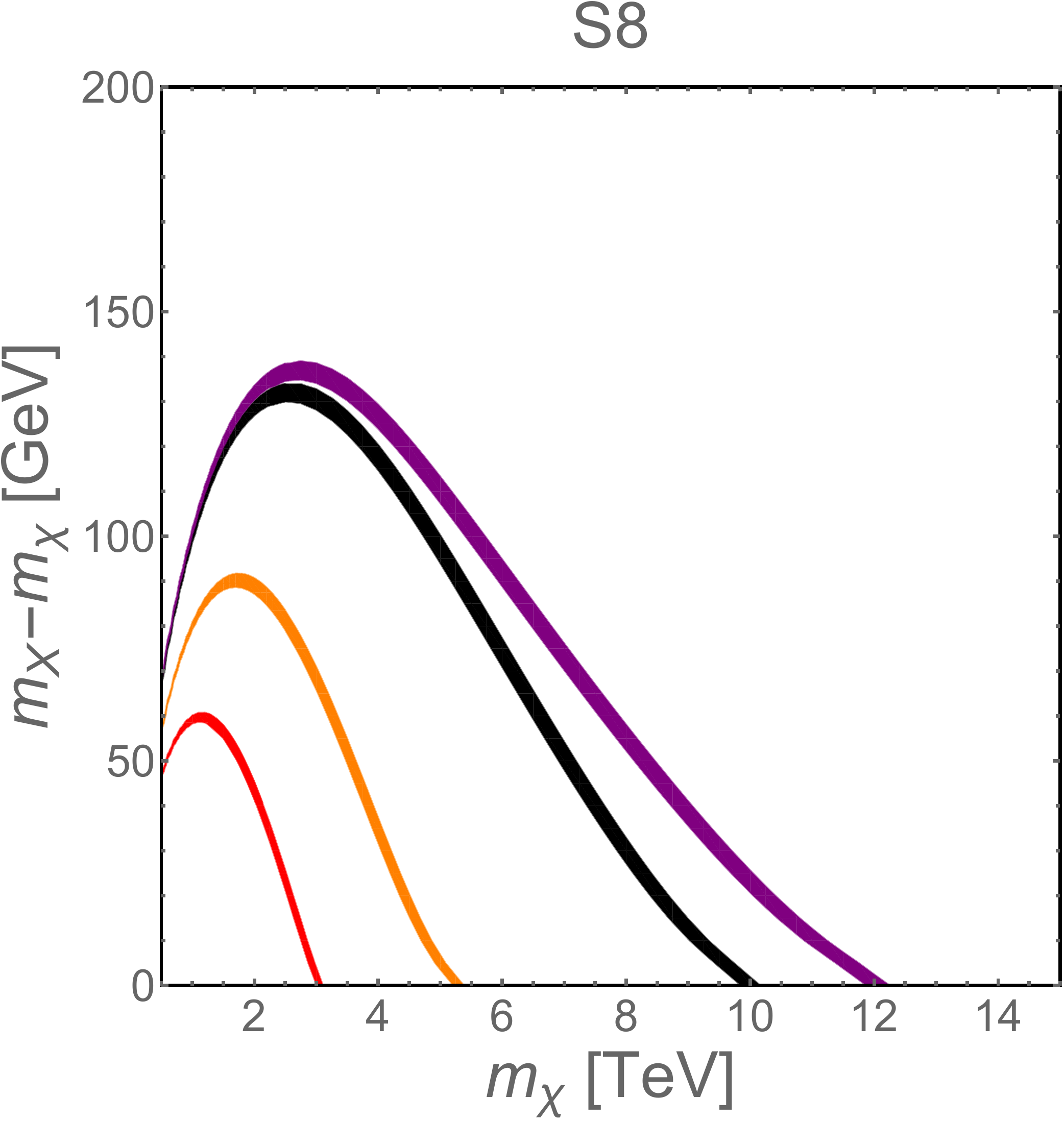} &
\hspace{-0.3cm}
\includegraphics[height=8cm]{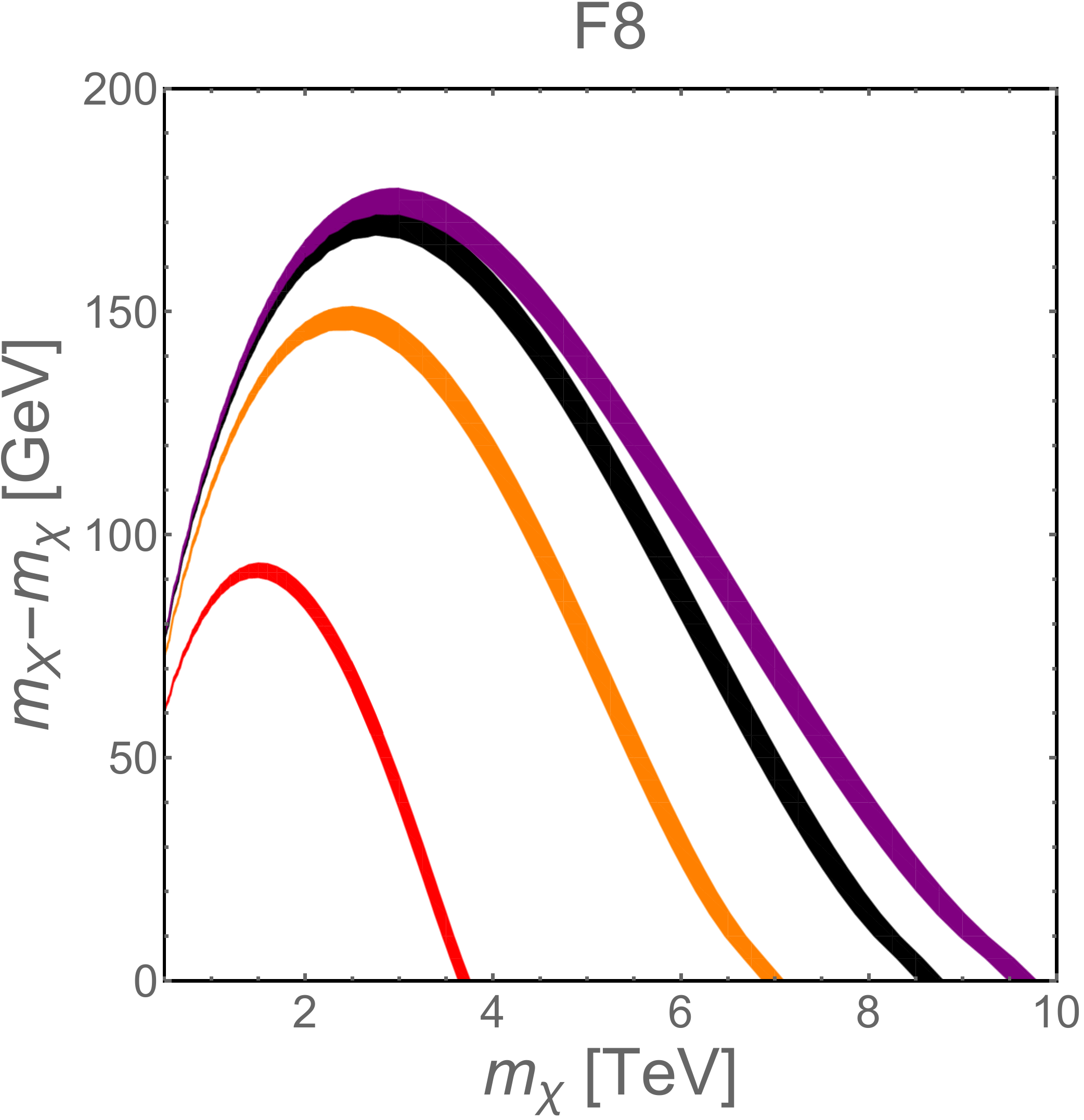} 
\end{tabular}
\end{center}   
\caption{\label{fig:mvsdmcountours}\it
The $(m_\chi, m_X - m_\chi)$ planes showing bands where $0.1151 < \Omega_\chi h^2 < 0.1235$ (3-$\sigma$ range of the Planck determination of the cold DM relic density),
for S3 (upper left), F3 (upper right), S8 (lower left) and F8 (lower right) coannihilating with a DM which has degrees of freedom $g_\chi = 2$. These results are
calculated without the Sommerfeld and bound-state effects (red), with the Sommerfeld effect but without bound-state effect (orange), with both the Sommerfeld and bound-state effects (black), and with the Sommerfeld effect and a factor of 2 enlargement of the bound-state effect (purple).
}
\end{figure}

\begin{figure}
\begin{center}
\begin{tabular}{c c}
\hspace{-0.3cm}
\includegraphics[height=5.5cm]{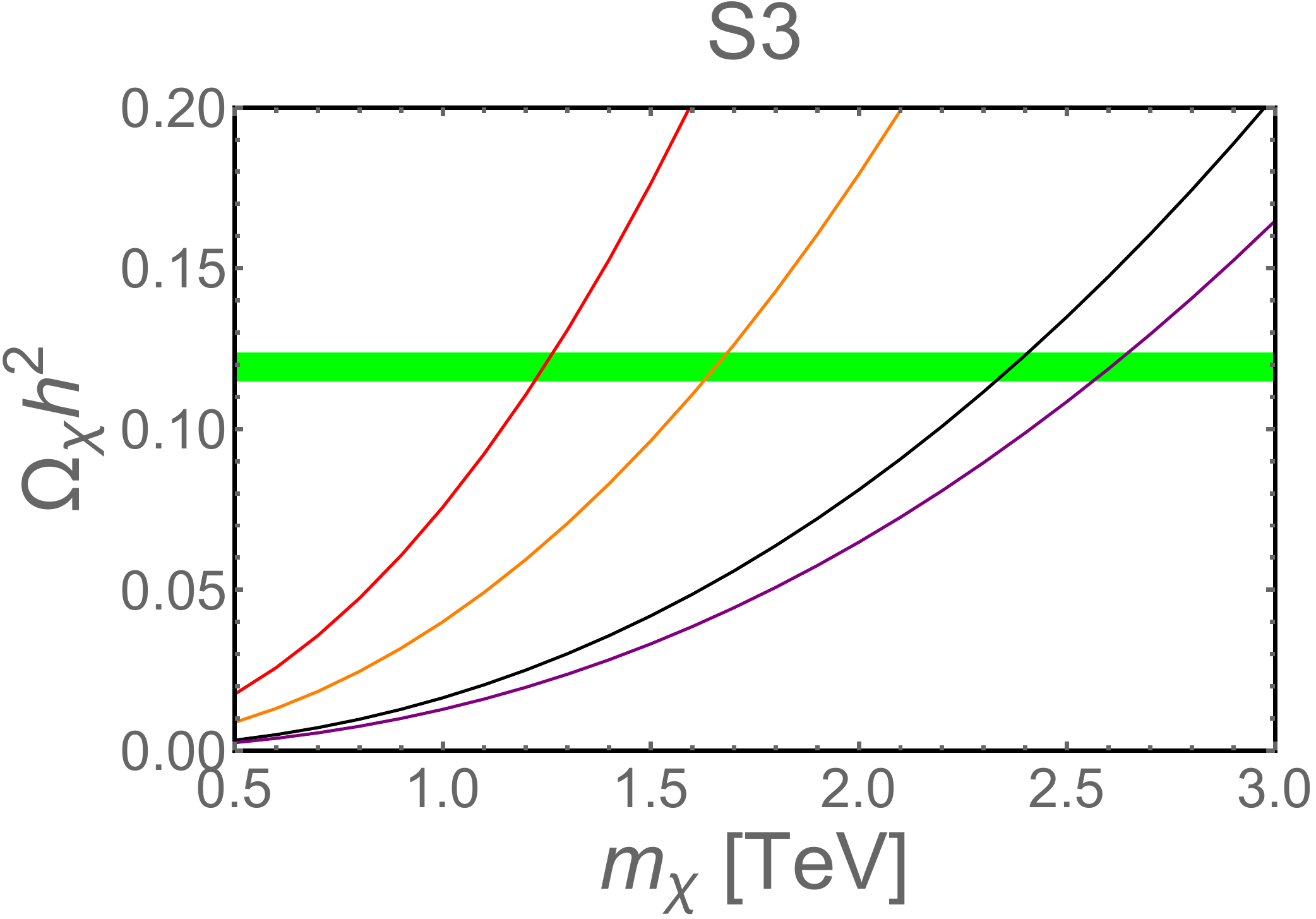} & 
\hspace{-0.3cm}
\includegraphics[height=5.5cm]{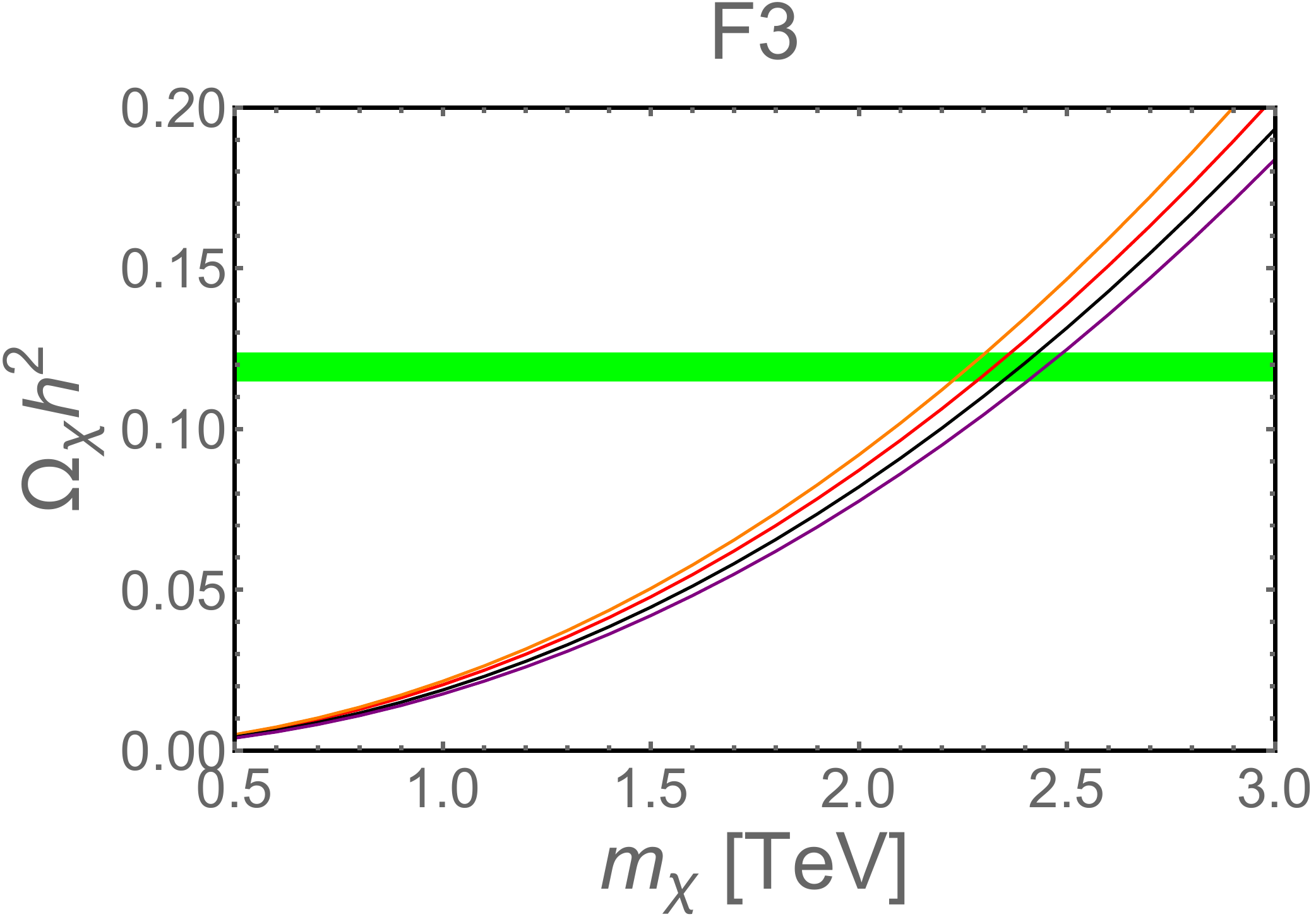} \\
\hspace{-0.3cm}
\includegraphics[height=5.5cm]{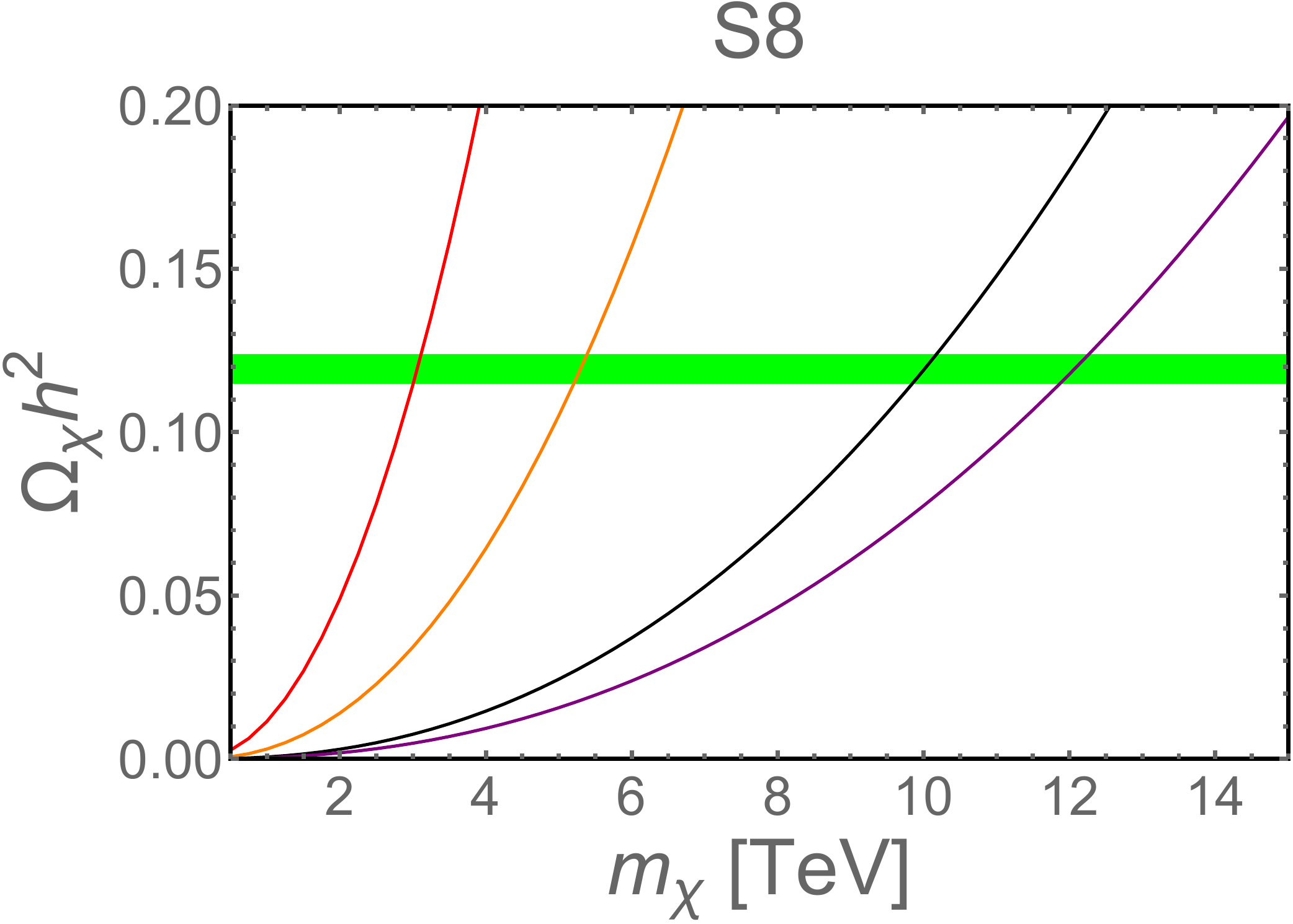} &
\hspace{-0.3cm}
\includegraphics[height=5.5cm]{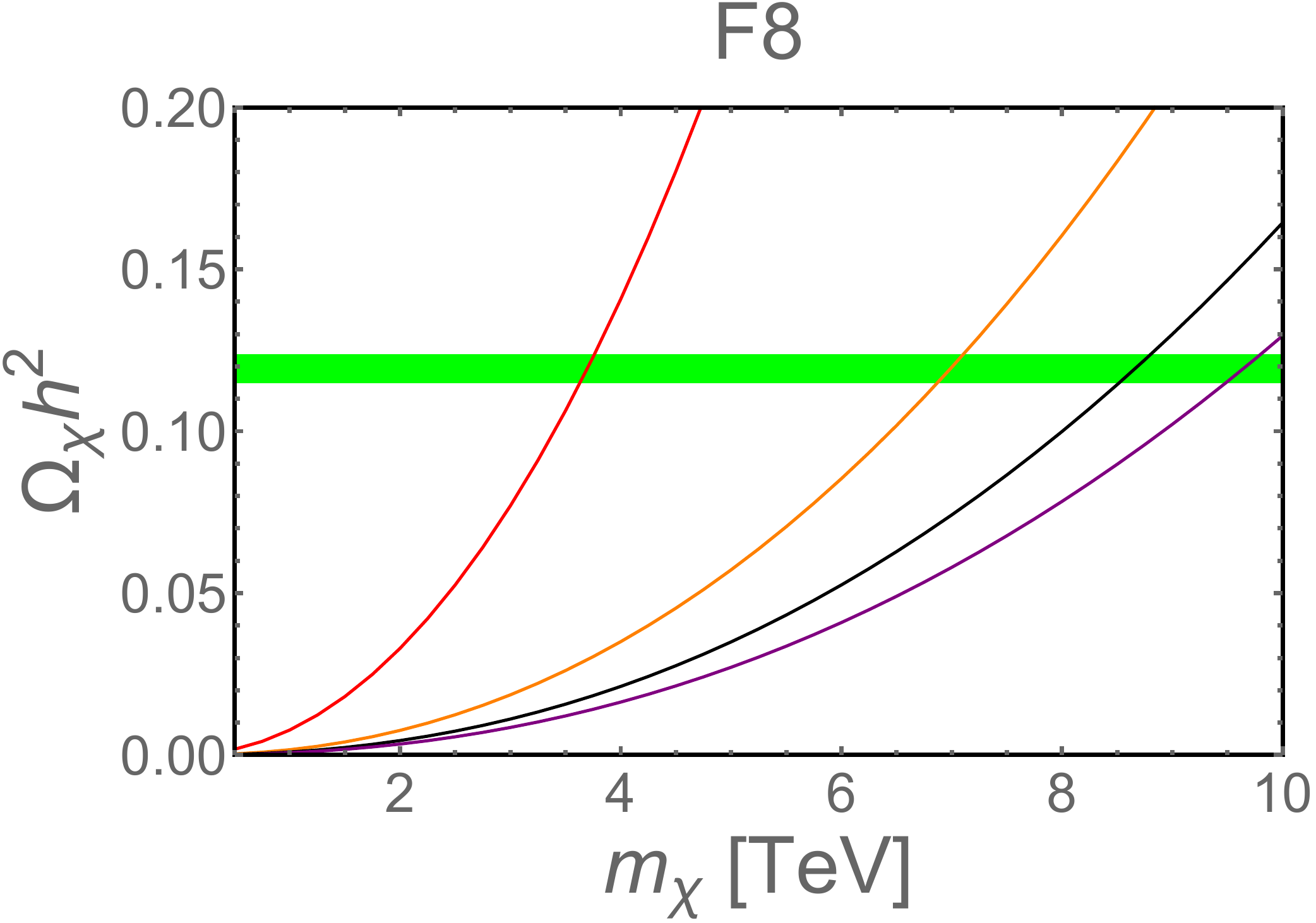}
\end{tabular}
\end{center}   
\caption{\label{fig:mvsohsq}\it
The locations of the endpoints (i.e., $m_X - m_\chi = 0$) of the coannihilation strips for different values of $\Omega_\chi h^2$, using the same color conventions as in Fig.~\protect\ref{fig:mvsdmcountours} for S3 (upper left), F3 (upper right), S8 (lower left) and F8 (lower right), respectively. 
The 3-$\sigma$ range $0.1151 < \Omega_\chi h^2 < 0.1235$ is shown by the horizontal green band. 
}
\end{figure}

Let us now study scenarios in which the WIMP DM $\chi$ has a mass close to a certain massive colored particle $X$, such that $\chi - X$ coannihilation is important in determining the DM relic abundance. We assume that the DM has degrees of freedom $g_\chi = 2$, corresponding to, for example, the Bino in the MSSM. The relic abundance of DM then depends only on the DM mass, $m_{\chi}$, and the mass splitting between DM and the colored particle, $m_X - m_\chi$. We plot in Fig.~\ref{fig:mvsdmcountours} in the $(m_\chi, m_X - m_\chi)$ planes the contour bands of the DM relic abundance falling within the 3-$\sigma$ range of the Planck determination of the cold DM density, $\Omega_{\text{CDM}} h^2 = 0.1193 \pm 0.0014$~\cite{1502.01589}. These bands are calculated using Eq.~(\ref{eq:bol}) without the Sommerfeld and bound-state effects (red), with the Sommerfeld effect but without bound-state effect (orange), with both the Sommerfeld and bound-state effect (black), and with Sommerfeld effect and a factor of 2 enlargement of the bound-state effect (purple). We can see that for the S3, S8 and F8 cases, on top of the Sommerfeld enhancement, bound-state effects further push upwards the largest mass splittings which can result in correct DM relic density. Also, the largest possible DM masses achieved at the endpoints of the coannihilation strips when the mass splittings approach zero, increase by $\sim 50\%, 100\%$ and $30\%$ with respect to the Sommerfeld-enhanced-only values, reaching $\sim 2.5, 11$ and 9 TeV for the S3, S8 and F8 cases, respectively~\footnote{The numerical differences for the F8 case in Figs.~\ref{fig:mvsdmcountours} and \ref{fig:mvsohsq} compared to Figs.~4 and 5 in~\cite{1503.07142} are due to a different use of $\alpha_s$ in the bound-state formation and dissociation cross sections, as well as the effect from the squark masses in the tree-level cross section of gluino pair annihilation into quark-antiquark pairs.}. For the F3 case in the upper right panel, however, the Sommerfeld and bound-state effects are much smaller compared to the other three cases. As can be seen from the positions of the red and orange bands, the Sommerfeld effect gives a slightly suppressed rather than an enhanced $\langle \sigma_{eff} v \rangle$~\footnote{The red and orange bands and curves in Figs.~\ref{fig:mvsdmcountours} and \ref{fig:mvsohsq} are consistent with the red and light green bands and curves in Figs.~1 and 2 in~\cite{deSimone:2014pda} for the S3, S8 and F8 cases. For the F3 case, the red band and curve presented here are also consistent with the ones in~\cite{deSimone:2014pda}, but the orange band and curve are different.}.

Fig.~\ref{fig:mvsohsq} shows in the ($m_\chi, \Omega_\chi h^2$) planes the locations of the endpoints of the coannihilation strips for different values of $\Omega_\chi h^2$, achieved when $m_X - m_\chi = 0$, for S3 (upper left), F3 (upper right), S8 (lower left) and F8 (lower right) coannihilating with a WIMP DM which has $g_\chi = 2$. The color conventions are the same as in Fig.~\ref{fig:mvsdmcountours}. The horizontal green band shows the 3-$\sigma$ range determined by Planck, $0.1151 < \Omega_\chi h^2 < 0.1235$. We can see that for the S3, S8 and F8 cases, for a given value of $m_\chi$ the Sommerfeld effect greatly reduces the calculated $\Omega_\chi h^2$ compared to the one without the inclusion of Sommerfeld factors. Also, the calculated $\Omega_\chi h^2$ is further significantly reduced after including the bound-state effect, in particular for a DM mass of TeV scale or larger. On the other hand, for the F3 case, again we see that the Sommerfeld and bound-state effects are small, and the Sommerfeld effect is opposite compared to the other three cases. 

\subsection{Implications of metastable particles on BBN and superWIMP DM abundance}
\label{sec:BBNandsuperWIMP}
It is well-known that the concordance of the standard BBN predications of the primordial light-element abundances with the values inferred from observational data, provides strong constraints on the abundance, mass, lifetime, and decay spectra of a massive particle decaying during or after BBN (see e.g.~\cite{astro-ph/0408426}). 
Since our focus in this work is to study the impacts of the bound-state effect of massive colored particles, we want to see how much the bound-state effect can change the BBN constraints on the  abundance and mass comparing to in particular the Sommerfeld effect, for a given lifetime and decay spectra which depend on other details of a specific particle theory model.   
With this in mind, we simply use the parametrization given in Eq.~(56) of~\cite{0807.0211} for the BBN constraints obtained by~\cite{astro-ph/0408426} for a massive metastable particle with a lifetime of $\sim 0.1-10^2$ sec and assuming that its hadronic decay branching ratio is 1 (which can be a good approximation for a massive colored particle), given as
\bea
Y_{X} \leqslant 1.0\times 10^{-13} \left(\frac{m_X}{1 {\rm \,TeV}}\right)^{-0.3} \;\;\; \text{for} \;\;\tau_X\sim 0.1-10^2 \; \text{s} \, .
\label{eq:BBNconstraint}
\eea
The above constraint comes from the would be overproduction of \he4, due to new proton $\leftrightarrow$ neutron interconversion reactions induced by the hadronic shower from X decays. In Eq.~(\ref{eq:BBNconstraint}), $Y_X$ is the sum of the yield for particle and anti-particle in our convention. 

\begin{figure}
\begin{center}
\begin{tabular}{c c}
\hspace{-0.3cm}
\includegraphics[height=5.5cm]{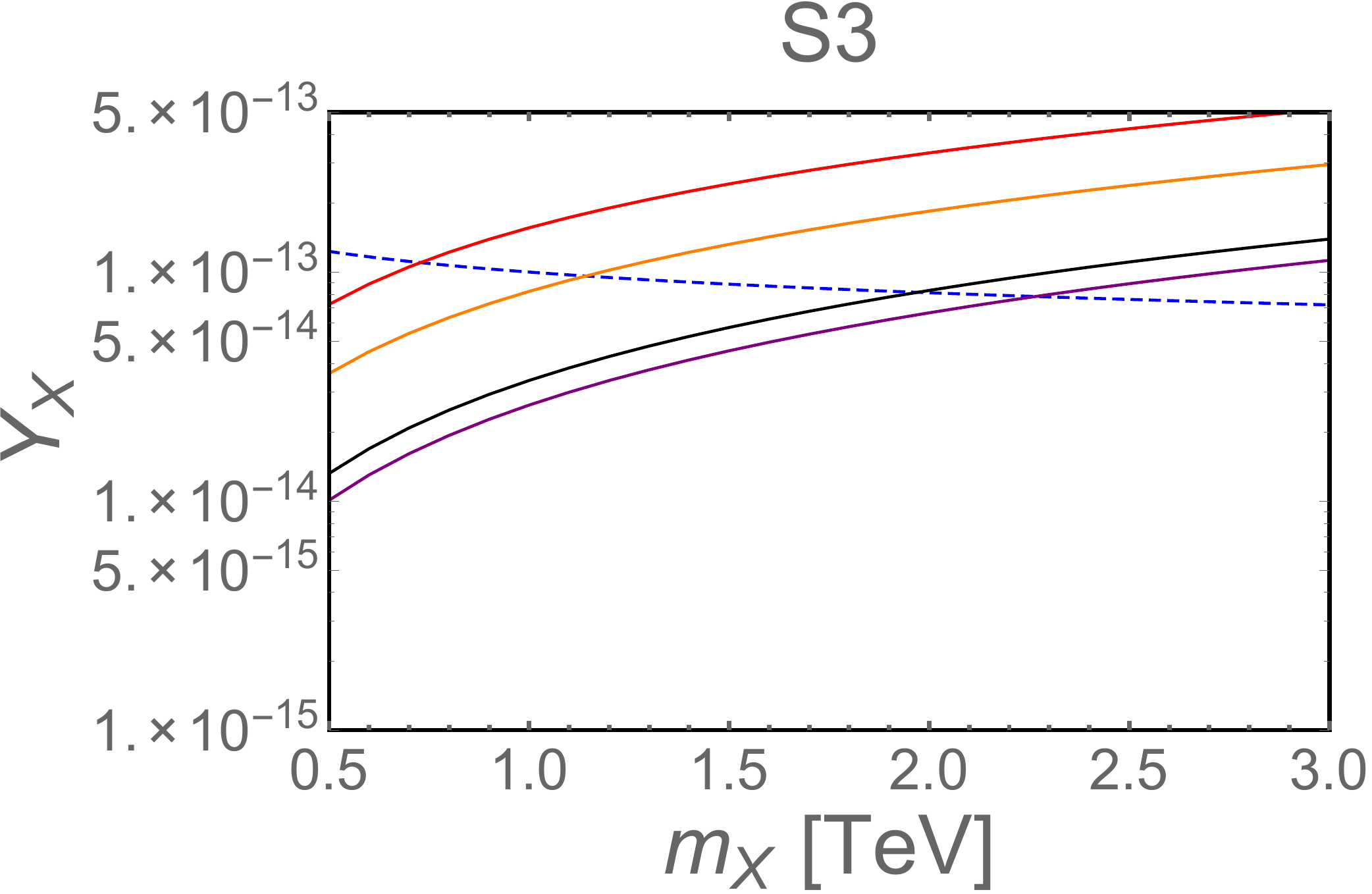} & 
\hspace{-0.3cm}
\includegraphics[height=5.5cm]{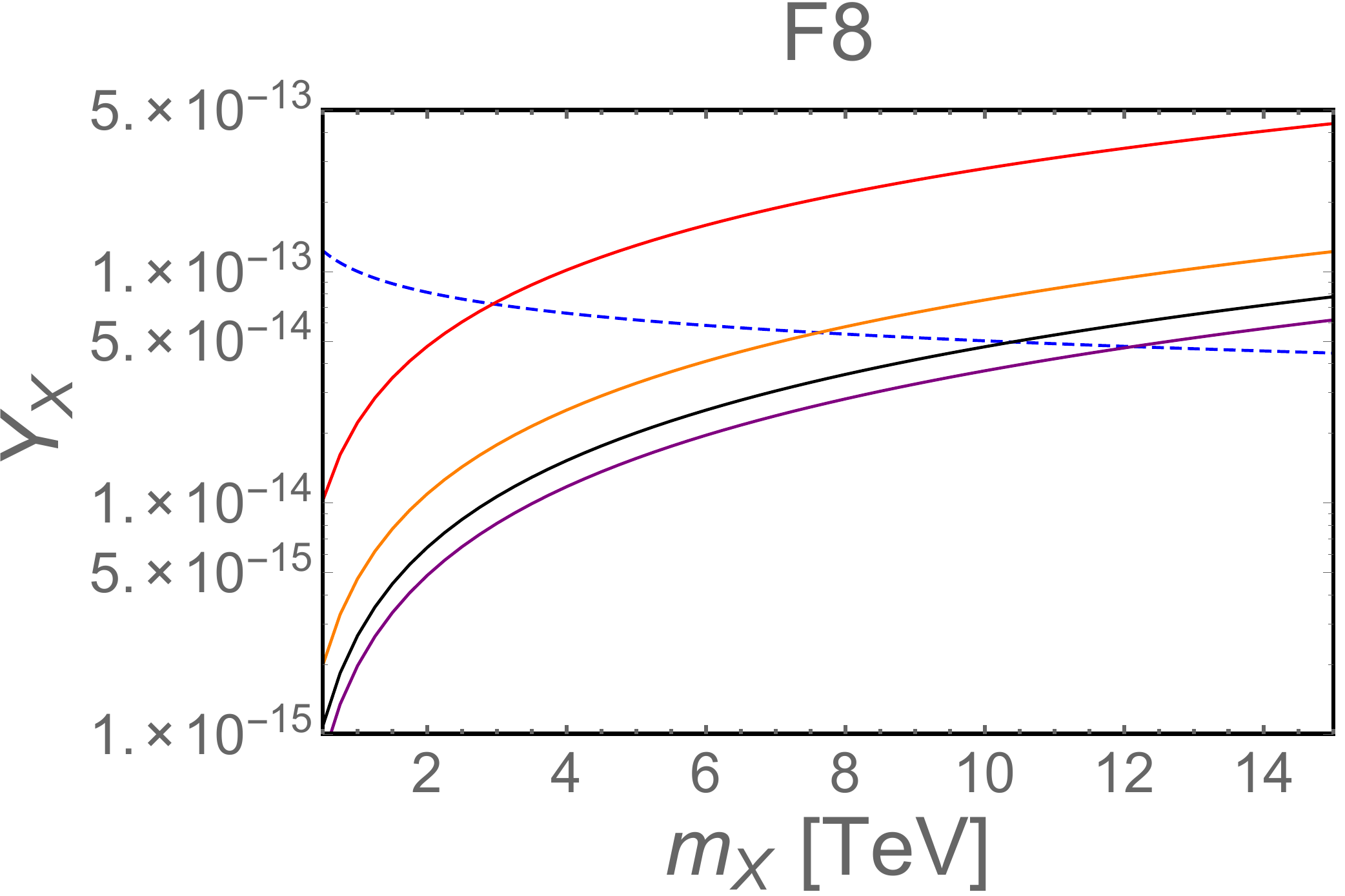} \\
\end{tabular}
\end{center}   
\caption{\label{fig:bbns3f8}\it
The total yield of the massive colored particles as a function of the mass, for the S3 (left panel) and F8 (right panel) cases, calculated without the Sommerfeld and bound-state effects (red line), with the Sommerfeld effect but without bound-state effect (orange line), with both the Sommerfeld and bound-state effects (black line), and with the Sommerfeld effect and a factor of 2 enlargement of the bound-state effect (purple line). The blue dashed line is the constraint given in Eq.~(\ref{eq:BBNconstraint}). 
}
\end{figure}

We show in Fig.~\ref{fig:bbns3f8} $Y_X$ as functions of $m_X$ for the S3 (left panel) and F8 (right panel) cases, calculated using Eq.~(\ref{eq:bol}) with the meanings of the variables understood as mentioned at the end of Sec.~\ref{sec:boltzmann} for massive colored particles. As before, the red, orange, black and purple lines are results without the Sommerfeld and bound-state effects, with the Sommerfeld effect but without bound-state effect, with both the Sommerfeld and bound-state effects, and with the Sommerfeld effect and a factor of 2 enlargement of the bound-state effect, respectively. The blue dashed line is given by Eq.~(\ref{eq:BBNconstraint}), and the parameter region above this line is excluded for an $X$ with a lifetime of $\sim 0.1-10^2$ sec. We can see that the bound-state effect pushes the allowed regions of $m_X$ to larger values compared to the ones with the Sommerfeld effect included only, namely, $\sim 1.1 \to 2.1$ TeV and $\sim 8 \to 11$ TeV for S3 and F8, respectively. As will be discussed more in the next section, since the LHC is pushing the exclusion limit of long-lived colored particles to TeV scale, it is useful to update the exclusion limit from BBN as well by including the previously omitted bound-state effect, so that we can be more confident to close or to leave open the mass window a long-lived colored particle can have. 

We now consider the contribution to superWIMP DM relic abundance from the out-of-equilibrium decays of the massive colored particles. The gravitino~\cite{hep-ph/0302215,hep-ph/0312262} and axino~\cite{hep-ph/9905212,hep-ph/0101009} LSP in $R$-parity conserving supersymmetric models serve as good examples of superWIMP DM. Using Eq.~(\ref{eq:relicsw}) and including the Sommerfeld and bound-state effects, the contribution can be parameterized approximately as
\bea
\Omega_{\text{SW}}^\text{non-th} h^2 &\sim& 0.1 	\left( \frac{m_{\text{SW}}}{1\,{\rm TeV}} \right) 
\left(\frac{m_{X}}{7\,\text{TeV}}\right)^{1.2}\, , \;\; \text{for} \; X={\rm S3} ,\\
\Omega_{\text{SW}}^\text{non-th} h^2 &\sim& 0.1 	\left( \frac{m_{\text{SW}}}{1\,{\rm TeV}} \right) 
\left(\frac{m_{X}}{60\,\text{TeV}}\right)^{1.2}\, , \;\; \text{for} \; X={\rm F8} .
\eea

We note that in our calculations of the relic abundance of massive colored particles, we have not included the possible further reduction of the abundance due to annihilations of heavy exotic color-neutral hadrons, which are formed by not-yet-decayed massive colored particles together with quarks and gluons after the quark-hadron phase transition in the early Universe~\cite{hep-ph/0611322,0811.1119}. In this sense, the BBN constraints given here are conservative. The contribution to $\Omega_{\text{SW}}^\text{non-th} h^2$ from the massive colored particle decays becomes smaller as well if there is further reduction of $\Omega_X h^2$ after the quark-hadron phase transition. 

\subsection{Electric charge corrections for the bound-state effect}
So far we have focused on bound-state effects with a gluon being emitted/absorbed in the bound-state formation/dissociation process. If a massive colored particle also carries some electric charge, for example, the squark in supersymmetry, a bound state can form (or be dissociated) by emitting (or absorbing) a photon, i.e., $X_1X_2\leftrightarrow \eta \gamma$. Also, the previously calculated bound-state formation/dissociation cross sections associated with gluon emission/absorption are modified due to the change of the potentials between the massive colored particles. To see the impacts of the electric charge on the bound-state effect, we use the S3 case as an example by assigning a charge $Q$ ($-Q$) to $S3$ ($\bar{S3}$), and consider the processes $S3\bar{S3}\leftrightarrow \eta g$ and $S3\bar{S3}\leftrightarrow \eta \gamma$. 

For $S3\bar{S3}\leftrightarrow \eta g$, we still consider the transition between the ($\text{color-octet}, L=1, S=0$) free pair state and the ($\text{color-singlet}, L=0, S=0$) bound state. By modifying the coefficients of the Coulomb potentials, $\zeta \to \zeta + \alpha_{\rm EM} Q^2$ and $\zeta' \to \zeta' + \alpha_{\rm EM} Q^2$, where $\alpha_{\rm EM}$ is the electromagnetic fine structure constant, the formulae given in Sec.~\ref{sec:formalism} still apply. Quantities depending on $\zeta$ and/or $\zeta'$, e.g., $a, E_B, \kappa$, and the cross sections and rates into which they enter, therefore all change. With an electric charge, for the bound state the previous attractive potential becomes more attractive. On the other hand, for the free pair state the previous repulsive potential becomes less repulsive, and even it can become attractive when $|Q|$ is large enough, i.e., $(-1/6) \alpha_s + \alpha_{\rm EM} Q^2$ is positive when $|Q| \gtrsim 3/2$. The bound-state annihilation decay rate changes as well due to the change of $\zeta$. 

For $S3\bar{S3}\leftrightarrow \eta \gamma$, the bound state and the free pair state are in the same color state, so that $\zeta = \zeta'$. Using dipole approximation~\footnote{We have checked that the dipole approximation is still justified with the inclusion of the electric charges we consider.}, we consider the transition between the ($\text{color-singlet}, L=1, S=0$) free pair state and the ($\text{color-singlet}, L=0, S=0$) bound state. The calculation is the same as for $S3\bar{S3}\leftrightarrow \eta g$, except that there is no color factors to worry about and the explicit coupling factor $\alpha_s$ in Eq.~(\ref{eq:ddis}) is changed to $\alpha_{\rm EM} Q^2$. The bound-state dissociation and formation cross sections are 
\bea
\sigma_{dis}^{\gamma} &=&  {2^{9} \pi^2 \over 3} \alpha_{\rm EM} Q^2 a^2 \left(\frac{E_B}{\omega}\right)^4 {e^{-4 \nu \arccot\nu} \over 1-e^{-2 \pi \nu}} \, , \\
\sigma_{bsf}^{\gamma} &=&  \frac{1 \times 2}{3 \times 3} \frac{\omega^2}{\left(\mu v_{rel}\right)^2} \times \sigma_{dis}^{\gamma}\, \, ,
\eea
where the superscript ``$\gamma$" indicates photon emission/absorption.
The quantities $a, E_B$ etc. are evaluated taking into account the change of potential due to the electric charge as mentioned above. In the thermally-averaged dissociation rate given in Eq.~(\ref{eq:disaverage}), $g_g$ is changed to $g_\gamma = 2$. The formula for the thermally-averaged formation cross section times relative velocity given in Eq.~(\ref{eq:thermalbsf}) stays the same. 

\begin{figure}
\begin{center}
\begin{tabular}{c c}
\hspace{-0.3cm}
\includegraphics[height=5.5cm]{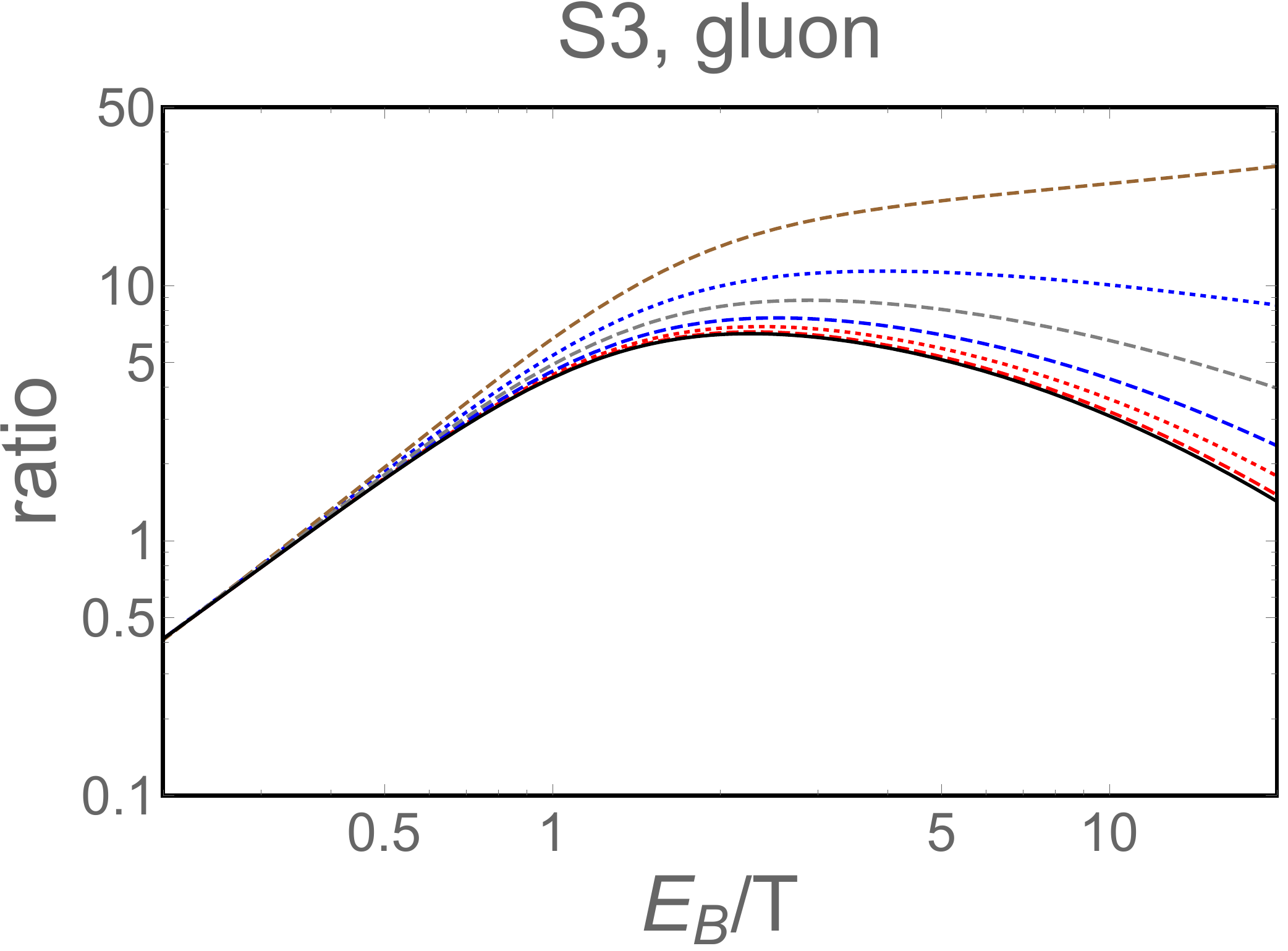} & 
\hspace{-0.3cm}
\includegraphics[height=5.5cm]{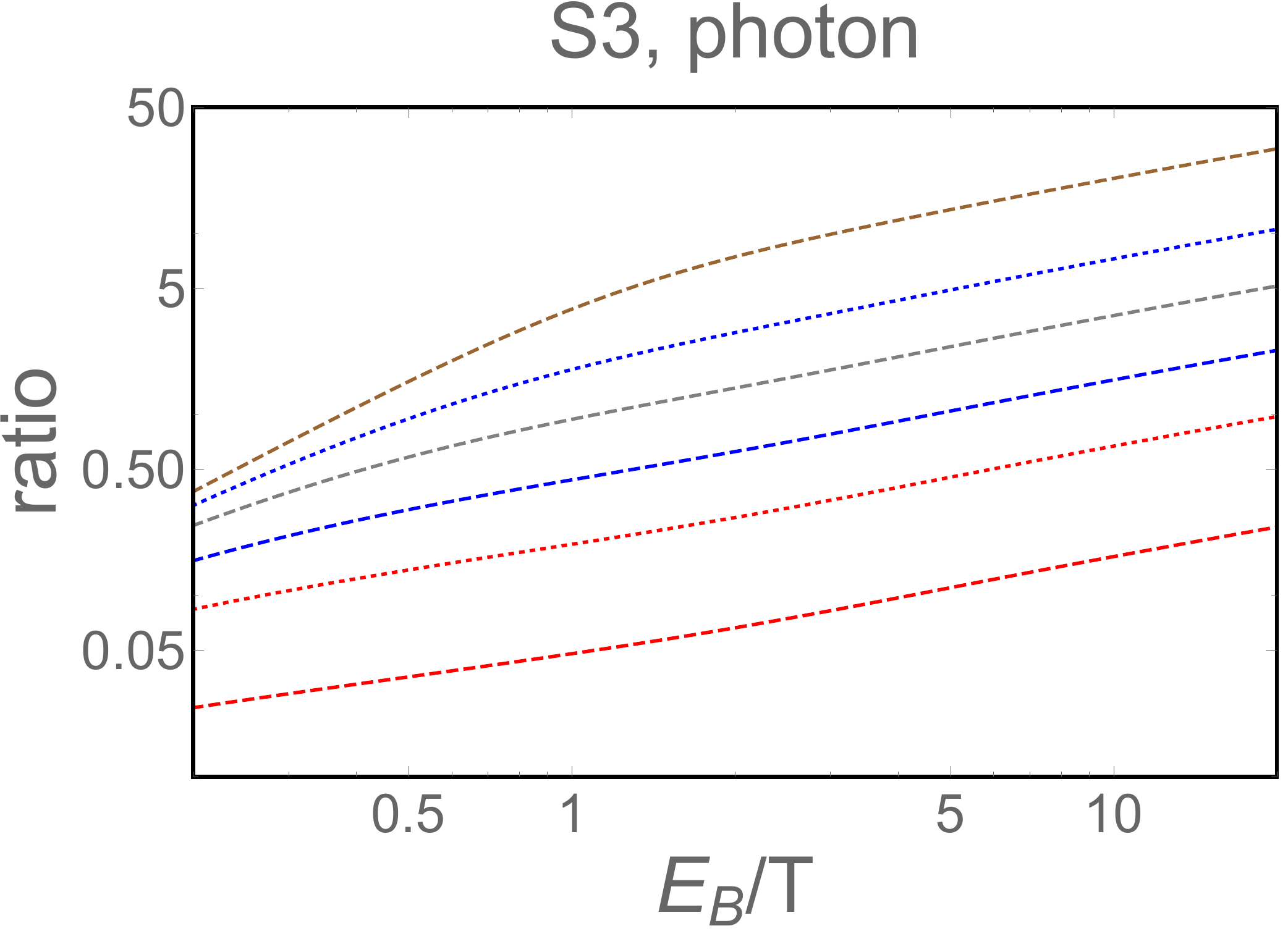} \\
\hspace{-0.3cm}
\includegraphics[height=5.5cm]{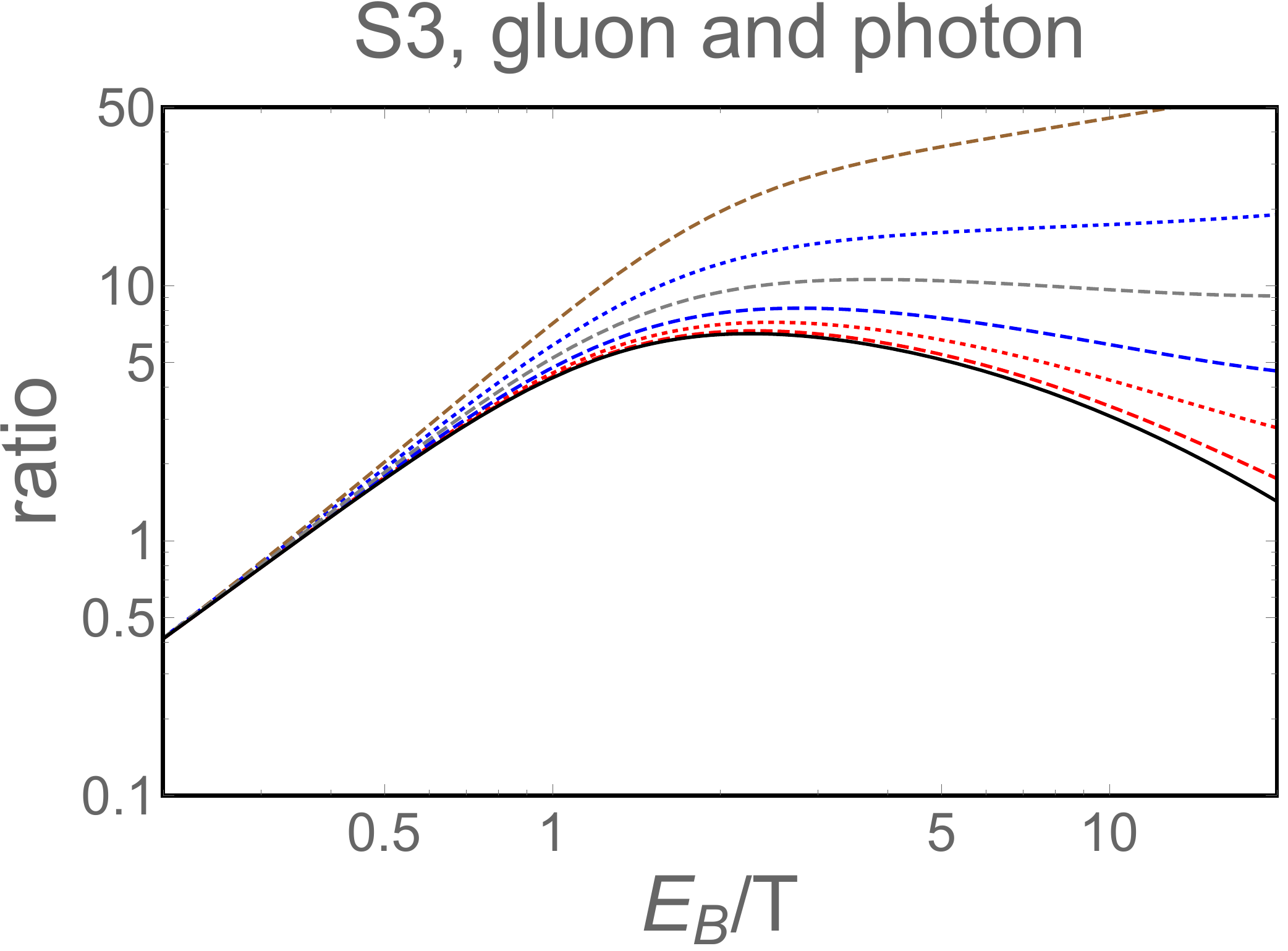} &
\hspace{-0.3cm}
\includegraphics[height=5.5cm]{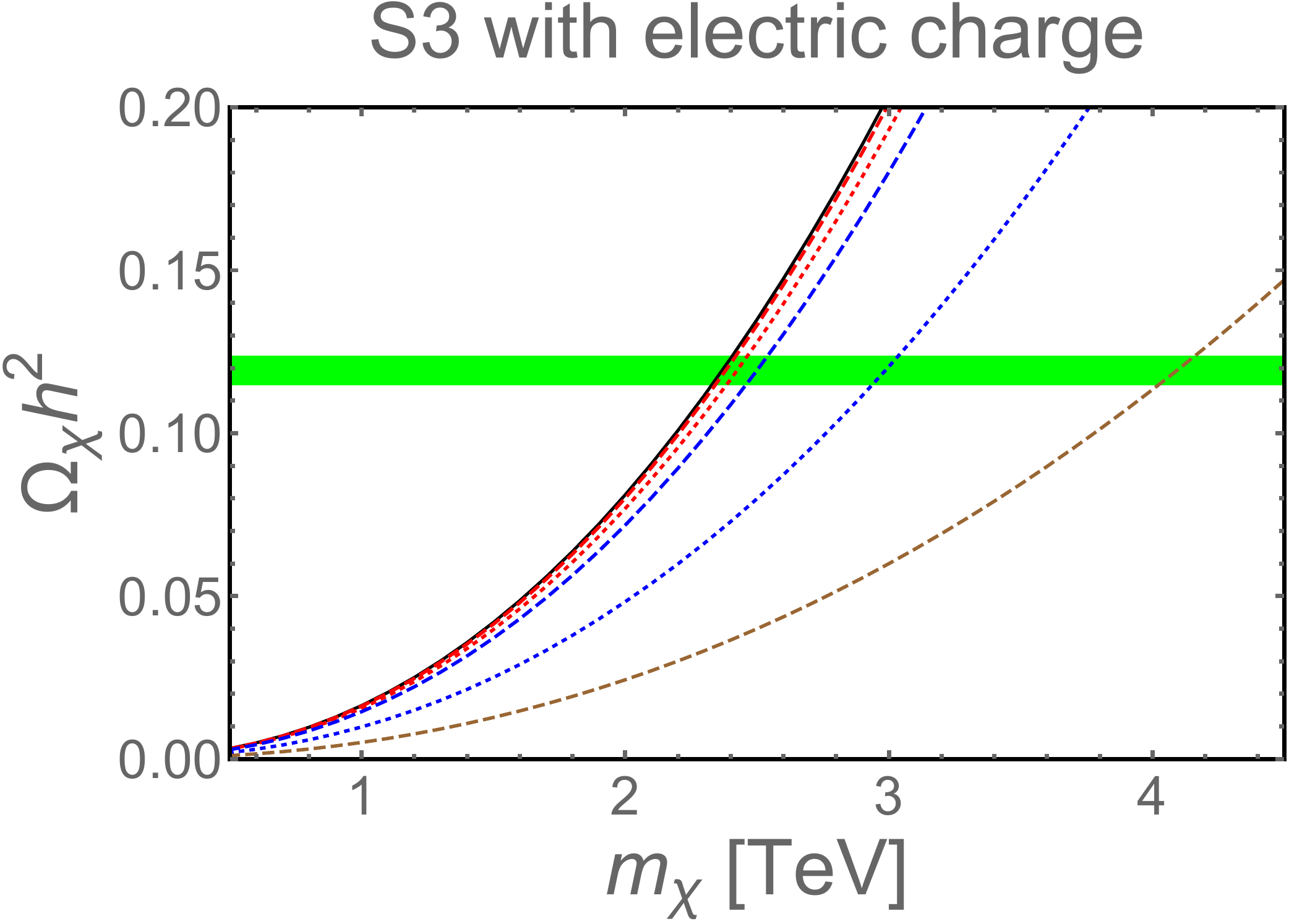}
\end{tabular}
\end{center}   
\caption{\label{fig:elecorrratio}\it
Impacts of the electric charge on the bound-state effect for the S3 case. The upper left panel is for the gluon emission/absorption bound-state effect, corresponding to Eq.~(\ref{eq:electricratiogluon}). The upper right panel is for the photon emission/absorption bound-state effect, corresponding to Eq.~(\ref{eq:electricratiophoton}). The lower left panel takes both of the above two into account, corresponding to Eq.~(\ref{eq:electricratioall}). The lower right panel shows the locations of the endpoints (i.e., $m_X - m_\chi = 0$) of the coannihilation strips for different values of $\Omega_\chi h^2$, for a WIMP DM with the degrees of freedom $g_\chi = 2$. The 3-$\sigma$ range $0.1151 < \Omega_\chi h^2 < 0.1235$ of the Planck determination of the cold DM relic density is shown by a horizontal green band. In all panels, the red dashed, red dotted, blue dashed, blue dotted and brown dashed lines correspond to cases of $|Q|=1/3, 2/3, 1, 2$ and $3$, respectively. The grey dashed lines in the upper left, upper right and lower left panels are for the case with the $Q$ chosen such that the potential for the free pair is zero. The black lines in the upper left, lower left and lower right panels are for the case of $Q = 0$. 
}
\end{figure}

We show in Fig.~\ref{fig:elecorrratio} the impacts of electric charge on the bound-state effect for the S3 case. The red dashed, red dotted, blue dashed, blue dotted and brown dashed lines correspond to cases of $|Q|=1/3, 2/3, 1, 2$ and $3$, respectively. The grey dashed lines in the upper left, upper right and lower left panels are for the case with the electric charge chosen such that $\zeta' = 0$. The black lines in the upper left and lower left panels are the same as the one in the left panel of Fig.~\ref{fig:ratio}, and the black line in the lower right panel is the same as the one in the upper left panel of Fig.~\ref{fig:mvsohsq}, all for $Q = 0$. We use $\alpha_{\rm EM} = 1/128$ in these plots, and take values of $\alpha_s$'s as noted in Sec.~\ref{sec:boundeffect}. The upper left panel shows the ratio
\be
\frac{\langle \sigma v \rangle_{bsf}^g \frac{\langle\Gamma \rangle_\eta}{\langle\Gamma \rangle_\eta+\langle\Gamma \rangle_{dis}^g}}{\langle \sigma v_{rel} (XX \to gg, q\bar{q})\rangle_{\text{w/o Sommerfeld}}} \,  \, ,
\label{eq:electricratiogluon}
\ee
where the superscript ``$g$'' indicates that only $S3\bar{S3}\leftrightarrow \eta g$ is considered. We can see that larger $|Q|$ makes the bound-state effect stronger. Also, for a large enough $|Q|$ such that the free pair potential becomes attractive (the blue dotted line and especially the brown dashed line), the behavior of the ratio at large $E_B/T$ becomes more like the black line in the right panel of Fig.~\ref{fig:ratio} where the free pair potential is attractive. The upper right panel shows the ratio
\be
\frac{\langle \sigma v \rangle_{bsf}^\gamma \frac{\langle\Gamma \rangle_\eta}{\langle\Gamma \rangle_\eta+\langle\Gamma \rangle_{dis}^\gamma}}{\langle \sigma v_{rel} (XX \to gg, q\bar{q})\rangle_{\text{w/o Sommerfeld}}} \,  \, .
\label{eq:electricratiophoton}
\ee
Again, larger $|Q|$ leads to larger bound-state effect with photon emission/absorption. By comparing with the corresponding lines in the upper left panel, we see that for $|Q| < 2$, at $E_B/T \lesssim 10$ the bound-state effect due to gluon emission/absorption is much larger than the one due to photon emission/absorption, while they become comparable for larger $|Q|$. The lower left panel shows the ratio
\be
\frac{\left(\langle \sigma v \rangle_{bsf}^g + \langle \sigma v \rangle_{bsf}^\gamma \right) \frac{\langle\Gamma \rangle_\eta}{\langle\Gamma \rangle_\eta+ \langle\Gamma \rangle_{dis}^g +\langle\Gamma \rangle_{dis}^\gamma}}{\langle \sigma v_{rel} (XX \to gg, q\bar{q})\rangle_{\text{w/o Sommerfeld}}} \,  \, ,
\label{eq:electricratioall}
\ee 
in which the numerator is the one entering into the Boltzmann equation as the second term in Eq.~(\ref{eq:xxtosm}). The shape of curves at smaller $E_B/T$ is controlled by the gluon emission/absorption bound-state effect, while the one from photon emission/absorption becomes important at larger $E_B/T$. 

We plot in the lower right panel the locations of the endpoints of the coannihilation strips for different values of $\Omega_\chi h^2$, after taking into account the total impacts of electric charge on the bound-state effects. As before, the horizontal green band shows the 3-$\sigma$ range of the Planck determination of the cold DM relic density, $0.1151 < \Omega_\chi h^2 < 0.1235$, and
we assume a WIMP DM with the degrees of freedom $g_\chi = 2$. Comparing to the black line which corresponds to $Q=0$, the $|Q| = 1/3$ and $2/3$ cases (corresponding to the charges of squarks) only slightly increase the endpoint values of $m_\chi$ for a given $\Omega_\chi h^2$, and by $\sim \mathcal{O} (10)$ GeV on the Planck band. For larger $|Q|$, the increase becomes significant, and the endpoint on the Planck band reaches $\sim 3$ (4) TeV for $|Q| = 2$ (3). 

\section{Collider constraints}
\label{sec:collider}
As shown in the previous section, the DM relic abundance in the WIMP DM coannihilation scenarios and the BBN constraints on the long-lived massive particle decays impose upper bounds on the masses of exotic  massive colored particles. On the other hand, collider experiments are constraining the masses from below. 

In the massive colored particle coannihilating with a WIMP DM scenario, the coannihilation region is characterized by a small mass splitting, which typically is difficult to probe at the LHC using the conventional multiple jets plus large missing energy searches. Massive colored particle pair production accompanied by a hard initial-state radiation, i.e., the monojet search, is utilized to constrain scenarios with such a compressed mass spectrum. For simplicity, we focus only on the $m_X - m_\chi \to 0$ region so that the kinematics of decay products of the colored particles can be ignored~\footnote{Opening the mass splitting $m_X - m_\chi$ will lower the monojet sensitivity as the missing energy is reduced by additional jets or objects from the decays of the massive colored particles. On the other hand, the signal efficiency of multi-jet plus missing energy searches is increased as the mass splitting opens, albeit the efficiency depends on how the colored particle decays. It is interesting to study how these complementary approaches can probe the coannihilation region. The detailed study is left for future work.}. The 13 TeV limits for S3 (stop) and F8 (gluino) are 0.32 TeV~\cite{1604.07773} and 0.63 TeV~\cite{1605.03814}, respectively. To impose limits on F3 and S8, we perform monojet simulations~\cite{ATLAS-CONF-2015-062,1604.07828} using the {\tt MadGraph-Pythia-Checkmate}~\cite{Alwall:2014hca,Alwall:2011uj,Sjostrand:2006za,1312.2591,1503.01123,deFavereau:2013fsa,Cacciari:2011ma,Cacciari:2005hq} pipeline. The obtained mass limits are 0.41 TeV and 0.43 TeV for F3 and S8, respectively. 
The results are summarized in Table~\ref{tab:dmlhc}, together with the endpoint values for the coannihilation strips including the Sommerfeld and bound-state effects read from Fig.~\ref{fig:mvsdmcountours}. 

Let us briefly comment on the discovery reach of a prospective 100 TeV proton-proton collider at an integrated luminosity of 3000 ${\rm fb}^{-1}$, in particular for the S3 case. The bound-state effect increases the mass range of DM significantly, making the coannihilation scenario more difficult to probe. With the inclusion of the bound-state effect so that the right-handed stop-Bino coannihilation strip ends at $\sim 2.5$ TeV when the Bino accounts for the total DM density (the inclusion of additional electroweak coannihilation channels~\cite{1501.03164} and a large lighter stop - heavier stop - Higgs coupling to heavier stop mass ratio~\cite{1404.5571} can shift the endpoint to even larger values), the ending part of the strip may be not within the discovery reach any more, though may be still within the exclusion reach given sufficiently low systematics~\cite{1404.0682}.

\begin{table}[t]
\centering
\begin{tabular}{l|cccc}
\hline
Bounds (TeV)& S3 & F3 & S8 & F8
\\ \hline
DM & 2.5 & 2.4 & 11 & 9 
\\ 
LHC & 0.32 & 0.41 & 0.43 & 0.63
\\ \hline
\end{tabular}
\caption{\it The DM masses at the endpoints of the coannihilation strips including the Sommerfeld and bound-state effects and giving the observed DM relic abundance, and the LHC monojet bounds for these coannihilation scenarios.}
\label{tab:dmlhc}
\end{table}

In the long-lived massive colored particle scenario, the produced massive colored particles at a collider form $R$-hadrons and travel through the detector with velocities significantly less than the speed of light, leaving ionization energy $dE/dx$ characteristically higher than that of charged SM particles. Searches for such events have been performed by both ATLAS and CMS collaborations at the LHC~\cite{1411.6795,1305.0491}. The current 13 TeV limits on long-lived (stable at collider scales) S3 (stop) and F8 (gluino) are 890 GeV and 1580 GeV, respectively~\cite{1606.05129}. While a dedicated simulation of $R$-hadron events at the LHC is beyond the scope of this work, in order to make a simple estimate of the bounds on long-lived F3 and S8, we assume that the signal efficiency and hadron formation probability of F3 (S8) equal to that of S3 (F8). The pair production cross section of F3 is estimated up to the next-to-next-to leading order (NNLO) using {\tt Hathor}~\cite{1007.1327}, while a next-to leading order (NLO) $K$-factor of 2 is used for the S8 scenario~\cite{1203.6358}. The obtained mass limits are 1.2 TeV and 1.4 TeV for F3 and S8, respectively. 
The results are summarized in Table~\ref{tab:bbnlhc}, together with the upper bounds from BBN for the masses of long-lived massive colored particles assuming lifetimes between 0.1 and 100 sec, including the Sommerfeld and bound-state effects, as discussed in Sec.~\ref{sec:BBNandsuperWIMP}. 

\begin{table}[t]
\centering
\begin{tabular}{l|cccc}
\hline
Bounds (TeV)& S3 & F3 & S8  & F8
\\ \hline
BBN & 2.1 & 1.7 & 17 & 11
\\ 
LHC & 0.89 & 1.2 & 1.4 & 1.6
\\ \hline
\end{tabular}
\caption{\it BBN upper bounds on the masses of long-lived massive colored particles assuming lifetimes between 0.1 and 100 sec, including the Sommerfeld and bound-state effects. Also shown are the LHC lower bounds on the masses of long-lived colored particles.}
\label{tab:bbnlhc}
\end{table}

\section{Summary}
\label{sec:conclusion}
We have studied in this paper the bound-state effects of exotic massive colored particles on DM relic abundance calculations in scenarios where the massive colored particles coannihilating with a WIMP DM. In general the bound-state effect increases the effective annihilation cross section through the formation and then annihilation decays of bound states, draining the number of DM particles in the thermal bath, when the massive colored particles and DM share the same discrete symmetry which stabilizes the latter, and provided the interconversion rate between the two particle species is fast enough compared to the Hubble expansion rate. For a given DM relic abundance, this effect allows a larger DM mass and a larger mass splitting between the massive colored particle and the DM. As examples, we consider the massive colored particles being complex scalars (S3) or Dirac fermions (F3) in the color $\text{SU(3)}$ fundamental representation, and real scalars (S8) or Majorana fermions (F8) in the adjoint representation. We find that the bound-state effect significantly increases the largest possible DM masses which can give the observed DM relic abundance, reaching $\sim 2.5, 11$ and 9 TeV for the S3, S8 and F8 cases, respectively. Comparing to the corresponding ones when considering only the Sommerfeld effect but without the bound-state effect, these values increase by $\sim 50\%, 100\%$ and $30\%$, respectively. The increase for the F3 case is smaller, but still the bound-state effect can more than counterbalance the Sommerfeld effect which is a suppression rather than an enhancement in this case.

We note that while the potentials for the bound states are attractive, due to color charge conservation the potential for an incoming massive colored particle pair can be attractive, zero or repulsive. In the early Universe bound states can form when incoming pairs have sufficiently large relative velocities to overcome the repulsive potential. In particular, for the S3 case, we find that although fading at low temperatures, the large bound-state formation cross section achieved when temperatures are comparable to the bound-state binding energy makes the bound-state effect significant enough, such that to probe the entire stop-Bino coannihilation strip can be quite challenging, if possible, even in a prospective 100 TeV proton-proton collider at an integrated luminosity of 3000 ${\rm fb}^{-1}$. 

We have also calculated the corrections for the bound-state effect when the massive colored particles carry electric charges. Using the S3 case as an example, we find that larger electric charge makes the bound-state effect stronger, and the enhancement can make the above mentioned $\sim 2.5$ TeV coannihilation endpoint to $\sim 3$ (4) TeV for $|Q| = 2$ (3), for which the incoming pair potential changes from being repulsive to attractive. However, for $|Q| < 1$, the enhancement is quite small. 

As we have briefly discussed, bound-state effects should also be included in calculations of superWIMP DM relic density from the decays of metastable massive colored particles, as well as when applying BBN constraints on long-lived massive colored particles. Furthermore, considering that BBN constraints and the DM relic abundance in coannihilation scenarios impose upper bounds on the masses of massive colored particles, we have studied the collider limits on the exotic massive colored particles we consider in this paper.

Before we close, we note that many other QCD bound states are possible. For example, in supersymmetry there can be di-squark and squark-gluino bound states~\cite{0912.0526}. Also, a squark and an antisquark with different flavors can form a bound state. Studying the effects of these bound states in specific supersymmetric models is left for future works. 

\section*{Acknowledgments}
We thank John Ellis, Wei-Chih Huang, Keith Olive and Satoshi Shirai for helpful discussions. F.L. thanks the hospitality of the University of G$\ddot{\rm o}$ttingen where part of this work was carried out. S.P.L is supported by JSPS Research Fellowships for Young Scientists and the Program for Leading Graduate Schools, MEXT, Japan. F.L. is supported by World Premier International Research Center Initiative (WPI), MEXT, Japan.

\appendix

\section{Thermally-averaged annihilation cross sections for S3, F3, S8 and F8}
\label{app:mat}
We follow the procedure given in~\cite{Srednicki:1988ce,hep-ph/9905481} in calculating the thermally-averaged $s$- and $p$-wave annihilation cross sections for a pair of massive colored particles into $gg$ or $q\bar{q}$, given as $\langle \sigma v_{rel} \rangle = a + b \, T/m_X+ \mathcal{O}\left((T/m_X)^{2}\right)$.  
Keeping results at leading order in $\alpha_s$, for $q \bar{q}$ final state we only need to evaluate the s-channel gluon exchange diagram, while for the $gg$ final state we need to evaluate the s-channel gluon exchange, t- and u-channel $X$ exchange diagrams, and the point interaction diagram when $X$ is a scalar. For the S3 case, the relevant squared amplitudes are covered in the stop - antistop annihilation calculations, given in~\cite{hep-ph/0112113}. The results for the F8 case are covered in the gluino - gluino annihilation calculations, listed in~\cite{1503.07142}. The results for the F3 case are covered in the heavy quark - antiquark annihilation calculations in the SM. For the relatively less familiar S8 case, the relevant part of the Lagrangian is $\mathcal{L} = \frac{1}{2} D_\mu \phi^b D^\mu \phi^b - \frac{1}{2} m_X^2 \phi^b \phi^b$~\cite{1412.5589}, with the color index $b$ summed over through 1 to 8. $\phi^b$ is a real scalar filed in the color $\text{SU(3)}$ adjoint representation, and $D_\mu \phi^b = \partial_\mu \phi^b + g_s f^{abc} G_\mu^a \phi^c$, where $G_\mu^a$ is the gluon field. 

Up to the common factor $\pi \alpha_s^2/m_X^2$, the $a$ and $b$ terms in $\langle \sigma v_{rel} \rangle$ we found are listed in Table~\ref{tab:sandpwave}. The results for the $q\bar{q}$ channel have summed over all 6 types of SM quarks, and we have dropped quark mass dependent terms, as we are considering massive colored particles much heavier than the SM quarks. 

\begin{table}[t]
\centering
\begin{tabular}{l|cccccccc}
\hline
&S3&F3&S8&F8
\\ \hline
$a \;\; \text{for} \;\; gg$           & $14/27$  & $7/27$  & $27/16$  & $27/32$
\\ 
$b \;\; \text{for} \;\; gg$           & $-61/27$ & $1/6$    & -261/32   & $9/64$
\\ 
$a \;\; \text{for} \;\; q \bar{q}$ &     $0$      & $4/3$    &    $0$      & $9/8$
\\ 
$b \;\; \text{for} \;\; q \bar{q}$ &  $4/3$      & $-14/3$ & $9/8$      & $-63/16$
\\ \hline
\end{tabular}
\caption{\it The coefficients a and b in $\langle \sigma v_{rel} \rangle$ for massive colored particle pair annihilation to $gg$ or $q\bar{q}$, up to the common factor $\pi \alpha_s^2/m_X^2$, for the S3, F3, S8 and F8 cases.}
\label{tab:sandpwave}
\end{table}

\bibliographystyle{aps}
\bibliography{ref}

\end{document}